\documentclass[11pt]{article}

\usepackage[T1]{fontenc}
\usepackage[utf8]{inputenc}
\usepackage[english]{babel}
\usepackage[margin=2.5cm]{geometry}
\usepackage{amsmath,amssymb,amsfonts,amsthm}
\usepackage{graphicx}
\usepackage{xcolor}
\usepackage{mathrsfs}
\usepackage{dsfont}
\usepackage{fancyhdr}
\usepackage[toc,page]{appendix}

\usepackage{subcaption}

\usepackage{newtxtext,newtxmath}
\usepackage{array}
\usepackage{booktabs}
\usepackage{multirow}

\usepackage{arydshln}
\usepackage[export]{adjustbox}
\usepackage{pdflscape}
\usepackage{bm}

\makeatletter
\@ifpackageloaded{hyperref}{}{\let\H@refstepcounter\refstepcounter}
\makeatother
\usepackage{hyperref} 

\theoremstyle{plain}
\newtheorem{thm}{Theorem}[section]

\newtheorem{lem}[thm]{Lemma}

\theoremstyle{definition}
\newtheorem{defn}{Definition}[section]
\theoremstyle{remark}
\newtheorem{Remark}{Remark}[section]

\title{Informative Risk Measures in the Banking Industry: A Proposal based on the Magnitude-Propensity Approach}
\author{
Michele Bonollo\thanks{Iason SRL, michele.bonollo@iasonltd.com and Politecnico di Milano, Italy.}\and 
Martino Grasselli\thanks{Department of Mathematics ''Tullio Levi Civita'', University of Padova, Italy, and De Vinci Research Center, Paris La D\'efense, France. martino.grasselli@unipd.it.}\and
Gianmarco Mori\thanks{Iason SRL, gianmarco.mori@iasonltd.com.}\and
Havva Nilsu Oz\thanks{Iason SRL, havva.nilsu@iasonltd.com. Acknowledgments: The authors thank Marco Veith for numerical support and Dr. Jorge Miguel Vegas (Intesa Bank) for valuable feedback.}
}
\date{} 

\begin{document}
\maketitle

\begin{abstract}

Despite decades of research in risk management, most of the literature has focused on scalar risk measures (like e.g. Value-at-Risk and Expected Shortfall). While such scalar measures provide compact and tractable summaries, they provide a poor informative value as they miss the intrinsic multivariate  nature of risk.
To contribute to a paradigmatic enhancement, and building on recent theoretical work by \cite{Faugeras21}, we propose a novel multivariate representation of risk that better reflects the structure of potential portfolio losses, while maintaining desirable properties of interpretability and analytical coherence. The proposed framework extends the classical frequency–severity approach and provides a more comprehensive characterization of extreme events. Several empirical applications based on real-world data demonstrate the feasibility, robustness and practical relevance of the methodology, suggesting its potential for both regulatory and managerial applications.

\end{abstract}



\section{Introduction and motivation}
The pursuit of effective risk measures in the financial industry is
an intriguing and long-standing endeavor that has garnered significant
attention from both scientific and regulatory perspectives. 
An adequate comprehensive risk management process typically consists
of the following key steps: (i) Identify risks, to detect all sources
of risk; (ii) Assess risks, i.e., measure them; (iii) Manage risks,
practically to avoid, transfer, hedge, or set limits and warnings;
(iv) Monitor and review risks, with ongoing activity. While some steps,
such as identification and management, relate to business knowledge
and the enterprise's organization and processes, the measurement step
(ii) is highly technical, involving several related challenging components,
such as: the choice of a "good" risk measure, the setup of its parameters,
a probabilistic model for random adverse events, the statistical calibration
of model parameters, and finally, the statistical estimation of risk
measures given empirical data. From a conceptual perspective, the
choice of risk measures, i.e., how to represent or summarize the randomness
of business variables (Profits and Losses in financial markets, Credit
Default events, Operational Losses), is a highly intriguing task,
as the adopted risk measures should satisfy certain quantitative and
qualitative properties. Briefly, given a univariate random variable
$X$ describing the potential profits and losses of a portfolio
(\textit{PnL} for brevity), one aims to summarize in a single scalar
value, the risk measure, the uncertainty that the bank (or insurance
companies) must manage. It is noteworthy that the risk evaluation
goal is often combined with return evaluation, as most capital allocation
choices are driven by the objective of optimizing the risk-versus-return
trade-off in some sense. Extensive work has been devoted in corporate
finance research to topics such as \textit{RORAC} (Return on Risk-adjusted
Capital) and \textit{RAROC} (risk-adjusted return on capital) indicators,
i.e., how to allocate capital, the scarce resource, by seeking the
best allocation on the efficient frontier concerning the enterprise's
risk appetite.  A seminal reference in this field, recently updated,
is \cite{brealey2019principles}. Adopting these models requires reliable
input in terms of (expected) return and risk estimation. While return
on an investment primarily poses statistical forecasting challenges,
capturing risk concisely for a portfolio is a more arduous task involving
subtle mathematical and conceptual points. We recall that risk, namely
the profits and losses behavior, is typically described
by a random variable. Furthermore, in many real-world models, the
distribution of this random variable is unknown, as it is the outcome
of a sophisticated model involving several risk factors, non-linear
relationships, non-Gaussian elementary distributions, and so on. This extends the problem from risk quantification to model uncertainty, where the distribution itself is not certain, usually known as \textit{model risk}, see \cite{Daniels2016} for a deep empirical survey. An extensive literature on risk measures exists, stemming
from both academic research and financial practitioners. The subtle
distinction between risk and uncertainty, along with a review of risk
measures, is provided in \cite{biglova2008desirable}.
The banking
and insurance sectors exhibit a high level of regulation; hence, regulators
(i.e., governments, central banks, banking authorities) have devoted
significant effort to defining appropriate risk measures to disclose
banks' risks to the market and all stakeholders. In particular, for over a decade now, international regulations and authorities have been emphasizing the concept of a \textit{Risk Appetite Framework} (RAF), which banks are required to establish, and the promotion of Risk Culture.
In this context, the concepts of \textit{Risk Capacity} (or \textit{Tolerance}) – the maximum risk that can be borne, especially \textit{Risk Appetite} – the desired risk level, and Risk Profile – the actual risk over time, are crucial, see see \cite{ecb2021targeted} (specifically the "Definitions" on page 15), and the Basel Committee's paper No. 328 for further guidance.
Risk culture pertains to the dissemination and awareness of risk among the bank's top figures (Board, Top Management), cascading throughout the organization, encompassing risk assumptions, risk types, models, and measures. Section 9 of the \cite{ecb2021targeted} guidelines provides insights on this aspect.
From a practical standpoint, the RAF is further articulated into a \textit{Risk Appetite Statement} (RAS), which contains high-level (strategic) indicators, known as Tier 1, for various risk types, followed by managerial indicators (Tier 2), and finally, operational or warning indicators at Tier 3 level. 
Of course, different regulations
prescribe different risk measures. The general ambitious goal of a
risk measure is to summarize the risky side of the uncertainty concept,
i.e., how much extreme losses could affect the expected return of
any investment strategy. As the extreme losses must be faced by the
own capital of the bank (or the insurance company), it is quite common to
say that the risks absorbe capital. At a very general level, we define
$X_{\theta}$ as the random variable describing the phenomenon
to be investigated, where $(\theta)$ represents the parameter (scalars
or arrays) characterizing the variable. Let  $F(x;\theta)$ 
be its cumulative distribution, i.e., $F(x;\theta)=P(X_{\theta}\leq x)$.
More generally, let ${\cal F}$ be the set of all random variables
on the real space (typically, ${\cal F}=L^{\infty}(\mathbb{R})$).
Following \cite{embrechts2005quantitative}, the ways to measure risk
can be grouped into four categories: the notional-amount approach,
sensitivity measures, risk measures based on scenarios (\textit{stress tests}), and
risk measures based on the \textit{PnLs} distribution. Focusing
on the last category, most portfolio
risk measures are statistical quantities describing the conditional
or unconditional loss distribution over some predetermined horizon.
The most popular include volatility $\sigma$, \textit{Value-at-Risk}
(VaR), and \textit{Expected Shortfall} (ES). More formally, a risk
measure is an application $\rho(X):{\cal F}\rightarrow\mathbb{R}_{+}$
mapping the random variable $X$ into a positive scalar value, representing
its uncertainty. A first point is to recall that risk relates to uncertainty,
but in most situations, banks focus on downside risk, i.e., the adverse
side of risk, where extreme losses in the business might occur with
a given frequency. For this reason, volatility $\sigma$ (estimated
standard deviation of portfolio returns), although widely popular
due to its simplicity and ubiquitous use in the asset management field
to rank products by their risk, is not adopted in financial regulation,
as it captures randomness from a symmetric perspective. 
To define what a good risk measure is, the notion of \textit{coherent}
risk measures has become prevalent, as outlined in the seminal paper
by \cite{artzner1997coherent}, see the next section for a detailed
overview. If we change perspective, we could wonder if the traditional
approach, i.e., seeking "the best" risk measure, is the proper approach,
or if an alternative strategy should be adopted. Can a one-dimensional
approach, i.e., a scalar measure summarizing adverse $PnLs$ results,
truly provide sufficient information about risk? Or should we move
to a multidimensional set of values to represent it?  To apply this perspective, which key theoretical or conceptual ideas need to be adapted? Are there alternative approaches to risk management beyond the search for an appropriate scalar risk measure? Describing risk by a
couple (or a few) indicators is quite common in fields other than
financial risk management. In the insurance sector, particularly in
the incident claims area, it is popular to summarize each event category
by two parameters: the frequency ($\lambda$) and the claim amount ($S$).
If the two are independent or approximately independent,
one can easily obtain the expected value of the claims to be
managed as $C=\lambda\cdot S$. The loss ($L$) in a given time horizon
($T$) is represented in the simplest case by a Compound Poisson process,
namely: $L(T)=\sum_{i:t(i)\leq T}S(i),$ where the sum of random losses
(lognormal, gamma, or other appropriate random variable) is extended
to a random number of events driven by the Poisson random variable.
A similar approach prevails in the operational risk sector in banks,
i.e., the risk of losses arising from incidents, errors, failures,
or fraud. In the Loss Distribution Approach
(LDA), a random number of negative events are recorded over
time, each with its severity. Even in the simplest lognormal approach
for severity, we do not know the exact closed-form distribution of
a sum of lognormal random variables, \textit{a fortiori} if the number
of events itself is random. To this extent, several analytical approximations
have been suggested, some from the mathematical finance field, see
e.g. \cite{turnbull1991quick}, others specifically developed in the
operational risk context, like e.g. \cite{peters2015single}. If we
move to enterprise risk management, even in large corporate contexts,
the methodology relies less on quantitative tools, as expert assessment
is the most popular approach. Once the risks in the company have been
identified and listed, for each risk, the process owner or the expert
panel is asked to complete a questionnaire to assign some Key Performance
Indicators (KPIs), such as: (i) The probability that no losses are
observed; (ii) The expected loss in normal situations; (iii) The magnitude
of losses in extreme cases, where "extreme" is qualified by some
percentile level or as the worst case over a very long horizon, etc.
A detailed guide to this approach is outlined in \cite{IOR2019}.
To summarize the above alternative approach and to avoid any confusion,
we point out that while the definition of a good (scalar) risk measure
focuses on the final output of uncertainty, i.e., the extreme losses
for the institution, the risk representation by a few parameters (probability
of the event, its severity, etc.) concerns the components, i.e., the
inputs to the final risk measure. The general idea underlying our
work is to combine these two aspects, namely the classical scalar
approach generally adopted in practice for defining a risk measure
in the risk management process, along with the multivariate perspective
based on its typical parameters on the input side.

From a theoretical standpoint, we refer to the recent work on the
magnitude-propensity approach in risk management by \cite{Faugeras21}.
The authors' groundbreaking work employs two closely related approaches
in their theoretical framework: optimal transport and quantization
techniques. Their methodology can be distilled into three key points.
Firstly, they argue that traditional single-valued risk measures, such as \textit{VaR} or \textit{ES}, are inadequate for fully capturing the underlying uncertainty.
Secondly,
they acknowledge the binary nature of losses - they either occur or
they don't. Lastly, they propose approximating the original risk model
(represented by the random variable $X$) with a binary variable,
which offers a more nuanced perspective on risk, balancing simplicity
with a richer representation of uncertainty. This simplified model
is characterized by two parameters: $p$, the probability of incurring
no losses, and $m$, the magnitude of losses when they do occur (with
probability $1-p$). Such distribution approximation problem is mathematically
formulated as mass transportation in Wasserstein metric space. The
optimal transportation to a discrete measure is analogous to the optimal
quantization problem, a well-established concept in Engineering and
Signal Processing literature. Consequently, the proposed approach
for quantifying risk on both magnitude and propensity scales can be
characterized as a specialized, constrained optimal quantization problem.
The formulation of \cite{Faugeras21} provides a rigorous mathematical
framework for risk assessment, bridging the gap between theoretical
optimal transport and practical risk quantification. It leverages
the Wasserstein metric's properties to capture the multidimensional
nature of risk, while the quantization aspect ensures computational
tractability and interpretability of the results. We immediately note that the approach of \cite{Faugeras21} differs from that of vector risk measures, first introduced by \cite{Jouini2004} and later revisited and developed by \cite{Lepinette2014}, \cite{Feinstein2015}, and \cite{Ararat24}, in order to account for proportional transaction costs in multi-asset markets.

Numerous attempts have been made to introduce additional properties aimed at improving these measures, identifying desirable characteristics, and clarifying what constitutes a well-defined notion of risk at the vector level. However, extending the same axioms used for scalar coherence often renders vector risk measures difficult to interpret and, in particular, to implement, since an additional selection procedure is typically required to determine a specific capital requirement or allocation rule within the set-valued framework, see \cite{Ararat24} for a recent overview of these methods.
The multidimensional approach we adopt here is of a different nature and not directly comparable to theirs.

Regarding our contribution to this field, from a theoretical point of view,
we endeavor to extend the framework of \cite{Faugeras21} to the 3-point
(i.e., zero, moderate, and extreme losses). In this multidimensional context, this contribution is significant. While the Risk Appetite Framework  encompasses not only market and credit risk (the focus of our examples) but also liquidity, operational, reputational, and other risks, one of the pressing issues for banks, as requested by the European Central Bank  during inspections, is "How did you determine the set of limits, i.e., the level of capacity/tolerance or warning levels?".
There is no common practice or guidelines on this matter. Most banks adopt the following approach: they consider the historical Value-at-Risk data, take the highest value, perhaps add a $10-20\%$ buffer, and set that as the tolerance level (keeping in mind that, at the bank level, as opposed to sub-portfolios, there may also be regulatory limits). In this regard, our optimal upper threshold $m_2$ corresponding to extreme losses could serve as an excellent "automatic machine" (recall that  ECB highly appreciates objectivity in the determination of the thresholds) to determine and support the setting of tolerance and/or capacity levels. For risks that involve only losses (operational, financial), the lower threshold $m_1$, corresponding to moderate losses, could also be an equally automatic tool for defining the risk appetite as observed in the past or as a "floor" for such appetite.
In the context of Risk Culture and top management's awareness of risks, a description using $(m_1, m_2)$ with the associated probabilities $( p_1, p_2)$ could provide an intuitive and complementary alternative to VaR and, more importantly, to the more complex Expected Shortfall, due to its familiarity with concepts such as severity and frequency.
Finally, for "financial conglomerates" under European directives, i.e., groups that include banks, insurance companies, and asset management firms, the regulations mandate the integration (effectively, uniformity and convergence) of measurement methods and risk culture across the various legal entities within the group. Our approach, which combines ideas from finance and insurance logic, could be a viable solution in this context.

Furthermore, our contribution
is also focused on applications, as we discuss certain constraints
that must be incorporated into the optimization problem to render
the solution more realistic in practical contexts. We apply
the technique to the historical simulation approach for \textit{VaR}
calculation and to the Montecarlo simulation for the Default Risk
Charge (DRC) measure, which are largely the most adopted by major
banks, precisely 19 out of 31 in the 2021 ECB review, see \cite{ecb2021targeted}.
To accomplish our objective, we utilize the \textit{PnLs} derived
from the real portfolio of a significant European bank. This makes
our work complementary to \cite{Faugeras21}, which primarily pertains
to the insurance sector, where we implement the third application case. The paper is organized as follows:
Section 2 reviews the relevant literature on risk measures; Section
3, after a brief background on quantization and the magnitude-propensity
approach, elucidates the extensions we implement to make this proposal
more suitable in the field of risk management; Section 4 illustrates
the application and the dataset employed for calibrating the models;
Section 5 contains our results, along with a discussion of the key
points; Section 6 summarizes the conclusions and outlines further
research avenues. We gather in the Appendix the technical proofs and some material on the numerical optimization procedures.

\section{Scalar Risk Measures: Review and Limitations}

Risk quantification and measurement constitute essential elements
in the domains of finance, insurance, and decision-making processes.
Throughout the years, a diverse array of risk measures has been proposed
and analyzed, each exhibiting distinct advantages and limitations.
Coherent risk measures represent a fundamental and well-established
framework for quantifying risk in financial and insurance contexts.
These measures adhere to a set of axioms that ensure a consistent
and rational approach to risk assessment. The coherence properties,
including subadditivity, monotonicity, translation invariance, and
positive homogeneity, provide a robust foundation for comparing and
aggregating risks across diverse portfolios and financial instruments,
see e.g. \cite{artzner1997coherent}. The field of risk measurement
has witnessed the development and application of numerous theories,
ranging from the widely utilized Value-at-Risk and Expected
Shortfall to \emph{Stress Tests}. Researchers and practitioners have
extensively explored these methodologies to evaluate potential losses
and manage uncertainties in various financial contexts. 
\textit{VaR}, a ubiquitously employed risk metric within the financial sector,
provides a probabilistic estimate of potential losses at a predetermined
confidence level over a specified temporal horizon. This measure enables
financial institutions to quantify and articulate downside risk with
precision, thereby facilitating informed decision-making and risk
management strategies. The diffusion of \textit{VaR} lies in its ability to
distill complex risk profiles into a single, comprehensible figure,
rendering it a useful tool for risk communication among stakeholders,
regulatory compliance, and internal risk control mechanisms. For a
given holding period and a level $\alpha\in(0,1)$ the \textit{VaR} of the
\textit{PnL} distribution $X$ is defined as 
\[
VaR_{\alpha}(X):=-\inf\{x\in\mathbb{R}:\mathbb{P}(X\leq x)>\alpha\},
\]
or, equivalently, $VaR_{\alpha}=\inf\{x\in\mathbb{R}:\mathbb{P}(Loss\geq x)\leq\alpha\}$.
One of the primary merits of \textit{VaR} lies in its simplicity and
interpretability, offering a lucid and succinct estimation of potential
financial losses. This characteristic enables risk managers to swiftly
evaluate and juxtapose risks across diverse portfolios or trading
positions. Nevertheless, \textit{VaR} is not without its limitations,
particularly in its sensitivity to extreme events, where its single-point
estimate may inadequately capture tail risk. The work of \cite{embrechts2005quantitative}
elucidates how \textit{VaR} embodies the propensity aspect of risk
by determining the leftmost quantile, yet fails to encapsulate the
magnitude of potential losses. Furthermore, \textit{VaR} has been
subject to substantial criticism due to its \textit{non-subadditivity}, a property
that contravenes the axioms of coherent risk measures. This deficiency
renders the aggregation of \textit{VaR} values across portfolios problematic.
To illustrate this point, one can readily construct an example utilizing
two sub-portfolios, $P_{1}$ and $P_{2}$, where the following inequality
holds: $VaR_{\alpha}(P=P_{1}\cup P_{2})>VaR_{\alpha}(P_{1})+VaR_{\alpha}(P_{2})$.
This mathematical representation underscores the inherent limitations
of \textit{VaR} in accurately reflecting the cumulative risk of combined
portfolios, thereby highlighting the need for more robust risk measurement
methodologies in complex financial environments. 

Conversely, the essential supremum, denoted as $\rho_{\infty}(X):=ess\sup X$,
represents the antithetical extreme to \textit{VaR}, quantifying the
maximal magnitude of a potential loss without estimating the associated
probabilities. This risk measure, also known as the worst-case scenario
or \textit{tail Value-at-Risk}, encapsulates the most extreme potential
loss, irrespective of its probability of occurrence. As such, it provides
a conservative estimate of the worst possible outcome but does not
offer a comprehensive risk assessment that incorporates both magnitude
and propensity aspects. One of the primary advantages of $\rho_{\infty}(X)$
lies in its simplicity and interpretability. By focusing solely on
the maximum potential loss, it enables risk managers to identify worst-case
scenarios and allocate capital reserves accordingly. In situations
where extreme events can have severe consequences, $\rho_{\infty}(X)$
provides a valuable upper bound for risk exposure, ensuring that institutions
are adequately prepared for the most adverse outcomes. However, this
extreme focus on maximum loss is not without drawbacks. By disregarding
probabilities, $\rho_{\infty}(X)$ neglects the likelihood of less
extreme but still significant losses. Consequently, risk managers
relying exclusively on this measure might overlook the impact of moderately
severe yet more probable events, potentially leading to suboptimal
risk management strategies. Furthermore, $\rho_{\infty}(X)$ exhibits
high sensitivity to outliers and extreme observations, rendering it
vulnerable to estimation errors and model mis-specifications. In practice,
financial data often display heavy-tailed distributions, implying
that extreme events occur more frequently than a Normal distribution
would suggest. As a result, $\rho_{\infty}(X)$ might overstate worst-case
scenarios, leading to excessively conservative risk assessments and
potential over-allocation of capital reserves. The theoretical underpinnings
and practical implications of the essential supremum in risk measurement
have been extensively explored in the seminal works of \cite{Rockafellar2002}
and \cite{Pichler2021}.

Expected Shortfall (\textit{ES}), also known as Conditional Value-at-Risk,
emerges as the most promising alternative to address \textit{VaR}'s
limitations. \textit{ES} quantifies the expected value of losses beyond the
\textit{VaR} threshold, providing information on extreme event severity
and tail behavior. It is formally defined as: 
\[
ES_{\alpha}=\mathbb{E}[Loss\vert Loss\geq VaR_{\alpha}].
\]
Expected Shortfall exemplifies a coherent risk measure, adhering to
the fundamental axioms of coherent measures (translation invariance,
subadditivity, positive homogeneity, and monotonicity, see \cite{artzner1997coherent}).
By quantifying potential losses beyond the \textit{VaR} threshold, 
\emph{ES} provides a more comprehensive and robust risk assessment,
particularly for heavy-tailed distributions, see \cite{acerbi2002coherence}.
This attribute renders \emph{ES} preferable in financial risk management,
where tail events can precipitate significant systemic consequences.
Regulatory bodies and financial institutions increasingly adopt \emph{ES},
acknowledging its capacity to address \textit{VaR}'s limitations and
more accurately represent extreme risk scenarios. Then \emph{ES} is
going to replace \textit{VaR} ($\alpha=99\%,h=10$ days) in the impending
Fundamental Review of the Trading Book (FRTB) regulation for market
risk minimum capital requirement calculations, see \cite{bcbs2019fundamental}.
In the forthcoming regulatory framework, Expected Shortfall is employed
to quantify market risk at a $\alpha=97.5\%$ confidence level and
different time horizons $h$ ($h$ = 10,20,40, 60, 120 days depending
on the liquidy ot the risk factors categories). However, financial
institutions are still required to validate their models using the
back-testing procedure based on \textit{VaR} calculated at both $\alpha=99\%$
and $\alpha=95\%$ confidence levels. This unconventional regulatory
setup stems from a decade-long debate surrounding the \textit{elicitability}
property of risk measures. Elicitability, derived from point forecasting,
requires statistical functionals to maintain consistency with their
evaluation through historical averaging, akin to $M-$functionals
(see \cite{Gneiting2011}). In this context, \textit{VaR} is elicitable
but not coherent, while Expected Shortfall is coherent and law-invariant
but not elicitable. Notably, ES exhibits joint elicitability with
\textit{VaR} (\cite{Fissler2016}). Furthermore, \emph{ES} possesses
the crucial property of backtestability, enabling rigorous performance
evaluation in historical simulations and real-world scenarios, as
elucidated by \cite{acerbi2017}, who underscore ES's capacity to
provide robust and reliable risk assessments, rendering it an invaluable
tool for risk managers in evaluating potential losses across diverse
financial scenarios\footnote{Our magnitude-propensity–based approach can be subjected to backtesting by extending the regulatory framework prescribed for VaR, see  \cite{Kupiec1995}, as well as  the more sophisticated joint VaR–ES testing procedures proposed by \cite{acerbi2017}. A comprehensive analysis of backtestability and the extension of the notion of elicitability to our multivariate framework requires  a dedicated investigation and will therefore be developed in a forthcoming paper.}.  More recently, the competition among the different
proposals for good risk measures has been enriched by some works about
the concept of \emph{observability} of a risk measure. Quite surprisingly,
the popular volatility indicator and the \emph{ES} show to
be more observable than \emph{VaR}. To this extent, an insightful
analysis is provided in \cite{acerbi2023}, by leveraging the concept
of \emph{sharp backtest}.

Finally, it is worth to note that in the FRTB regulation the banks
must also calculate another risk measure, the \emph{Default Risk Charge}
(DRC), namely a 1-Year \emph{VaR} with $\alpha=99.9\%$. This capital
requirement aims to capture the default risk in the trading book.
Until the 2008-2011 financial crisis, the so called \emph{market risk
}building block in the Basel regulation included only the \emph{spread
risk}, i.e. the price uncertainty coming form the spread level, e.g
in the bond evaluation. According to the Basel 2.5 Reform, it was
requested to estimate the potential losses coming from the default
(or migration) events also for the trading book of the bank, see \cite{IRC}.
While the $PnLs$ of the traditional market risk come from the movements
of the risk factors (equities, interest rates, forex rates, spreads,
commodities) that are commonly assumed to have continuous distribution,
in the credit risk  case the portfolio losses are originated by  the default binary events $D_{n}$ associated to the obligors ($n=1,\cdots,N$):
\[
Loss_{DRC}=\sum_{n}EAD_{n}\cdot\mathbb{\mathbf{1}}_{ D_{n} }\cdot LGD_{n},\label{DRC}
\]
where $EAD_{n}$ and $LGD_{n}$ represent, respectively, the amount
of exposure and the fraction that the bank is not able to recover
once the default happens, modeled as a deterministic coefficient in
the range $\left[0,1\right]$, or by a random variable, usually a \emph{Beta (a,b)}.
It is important to notice that the combinatorial effects associated with default events lead to an intrinsic instability of any risk measure, due to the fact that  extreme
values of the distribution are so pronounced that the VaR no longer provides an adequate representation of
tail risk. We will illustrate this effect in the Appendix \ref{AppendixCredit}. 

\textit{Stress tests} represent a pivotal tool for risk analysis,
enabling institutions to assess resilience against adverse market
conditions and extreme events. These tests simulate hypothetical crisis scenarios, such as
economic recessions, market shocks, or event-related crises, by subjecting portfolios or financial systems to
various stress factors.
By modeling extreme conditions, stress tests identify vulnerabilities
and quantify potential losses under adverse circumstances. The results
provide insights into an institution's risk exposure, capital adequacy,
and overall financial stability. This risk management tool has become
increasingly crucial for regulatory compliance and is widely adopted
by financial institutions to ensure robust risk management practices.
Notably, stress tests have been instrumental in enhancing the financial
sector's resilience following the 2008 global financial crisis. By
simulating scenario outcomes, stress tests empower institutions to
implement appropriate risk mitigation strategies and maintain financial
stability in turbulent times. From a conceptual standpoint, stress
tests diverge significantly from measures such as Value-at-Risk and
Expected Shortfall. On one hand, stress tests are more objective in
their results and are based on a simple "\textit{what-if}" logic: given a
scenario, banks can calculate the related \textit{PnL} with high accuracy.
On the other hand, regulations mandate the use of \textit{extreme
yet plausible} scenarios, a goal that is arduous to achieve even for
regulatory authorities. Furthermore, stress test results lack information
about the associated probabilities.

In this respect, 
the common practice consists of some heuristic methods. To determine thresholds for risk tolerance or risk appetite in the their risk appetite framework, typically some banks consider the past time average of the Value-at-Risk and increase it by some subjective quantity (e.g. $20\%$), other banks calculate directly the \textit{Stressed VaR}, i.e. \textit{VaR} under stressed parameters. Such approach is quite reasonable but exhibits some drawbacks, being too judgmental without a rational basis. Furthermore,  these methods do not add any relevant additional information about risk beyond what is provided by the \textit{VaR} itself. Additional information is required, which only a multidimensional approach can provide, as will be explained in the following section.
\textit{VaR} does not provide insights into the underlying risk factors or the potential severity and frequency of extreme events. By simply scaling the \textit{VaR} up or down by a fixed percentage, the resulting stressed scenarios fail to capture the complex dynamics and interdependencies of risk factors, as well as the potential for tail events that may deviate significantly from the assumed distribution.
A multidimensional approach, on the other hand, can incorporate additional risk dimensions, such as severity and frequency, as well as the potential for non-linear interactions between risk factors.
In the following section, we will explore a multidimensional approach that incorporates severity and frequency dimensions to provide a more comprehensive and informative framework for risk management and scenario analysis.

\section{ From Scalar to Vectorial Risk Measures and from Continuous to Discrete
Distributions}

\cite{Faugeras21} observe that \textit{VaR} and
\textit{ES} are traditionally calculated for popular confidence levels $\alpha$
(e.g., $95\%,99\%,97.5\%$), introducing a degree of subjectivity
to these risk measures and highlighting potential weaknesses in comparing
distinct \textit{PnLs} distributions. The magnitude-propensity approach
aims to address these limitations. Given the random variable for the
\textit{PnL}, say $X$, the approach consists in defining a discrete
binary random variable $Y$ that works as an informative summary of
the whole $X$, whose distribution is given by 
\[
P^{Y}=(1-p)\cdot\delta_{0}+p\cdot\delta_{m},
\]
where $p$ is the probability that the loss is zero, $m$ represents
the magnitude of the loss and $\delta_x$ denotes the Dirac distribution concentrated at the point $x$.
Various methodologies have
been developed to discretize continuous risk distributions, providing
decision-makers with robust tools for risk assessment and strategic
planning. 
 Two main approaches
emerge in the literature - the \textit{optimal transport}  approach and the
\textit{optimal quantization} approach.
The  \textit{optimal transport} method
approximates continuous distributions by minimizing mass transportation
costs between distributions, as explained by \cite{villani2009}.
On the other hand, \textit{optimal quantization}, described by \cite{graflush00}, \cite{PagPham2004}
and \cite{LusPages23}, represents risk data using finite point grids\footnote{
We also mention other methods like $K-$means clustering, studied by \cite{LiuPages2020}, which groups similar
risk data points, then revealing underlying risk patterns, and the  quantile-based
methods, explored by \cite{ChernHong2003}, that partition risk distributions
based on predefined quantiles, focusing on specific risk thresholds.}.

 While both the optimal transport
and quantization approaches aim to disentangle the magnitude and propensity
aspects of risk, they employ distinct mathematical frameworks to achieve
this goal. The optimal transport approach represents the search for
the binary random variable $Y$ as a mass transportation problem,
whereas the quantization approach treats it as an optimal discrete
representation problem. However, optimal transportation to discrete measures
also corresponds to a special case of optimal quantization, in fact
the proposed approach to quantify risk on both the magnitude and propensity
scales amounts to a special, constrained, optimal quantization problem.
Specifically, we briefly review the formulation of the bivariate magnitude-propensity
risk measure and its optimization using \textit{Wasserstein distance}
for the optimal transport approach. Furthermore, we explore the intricacies
of the optimal quantization problem and its relevance in approximating
risk distributions with discrete measures. We then attempt to establish
a solid theoretical foundation for our research, which aims at extending
the original 2-points distribution framework of \cite{Faugeras21}
by developing a novel 3-points distribution approach. Our extended
methodology incorporates an additional quantile, enabling a more informative
and comprehensive risk quantification process.

\subsection{Optimal Transport Approach}

The optimal transport (OT) problem, formally introduced by \cite{Kantorovich42},
addresses the transformation of a probability distribution $P^{X}$
of the random variable $X$ to a distribution $P^{Y}$ of the variable
$Y$, while minimizing the associated transition cost function $c(x,y):X\times Y\rightarrow[0,+\infty]$.
The Monge-Kantorovich optimal transportation problem has a clear physical
interpretation: considering the random variables as material locations,
then $c(x,y)$ represents the cost of transporting a unit of material
from $x$ to $y$. Optimal cost functions typically represent the
transport cost as the product of mass and inter-location distance.
The problem is constrained by complete material relocation. This framework
has diverse applications, spanning economics, image processing, and
machine learning, providing a robust methodology for distribution
transformation analysis and optimization. The formulation maintains
scientific rigor while offering concise elegance, emphasizing the
concept's broad academic and practical relevance. This version further
condenses the information while preserving the scientific tone and
elegant expression. It highlights the key points about cost function
formulation, problem constraints, and the wide-ranging applicability
of the optimal transport framework. Mathematically, the Monge-Kantorovich
optimal transport problem is formulated as: 
\[
\inf_{\pi\in\Pi(P^{X},P^{Y})}\int_{\mathbb{R}\times\mathbb{R}}c(x,y)\,\pi(dx,dy)
\]
where $\Pi(P^{X},P^{Y})$ is the set of all transport plans, i.e.,
joint probability measures on $\mathbb{R}\times\mathbb{R}$ with marginals
$P^{X}$ and $P^{Y}$, respectively. Under regularity conditions,
the optimal transportation plan is defined as a Monge map $T$, namely
$\pi(x,y)=\pi(x,T(x))$, see e.g. \cite{Rachev1998}, \cite{villani2009}.
In the degenerate case where $P^{Y}=\delta_{m}$, with $m\in\mathbb{R}$,
$\Pi(P^{X},P^{Y})$ reduces to the singleton product measure $\{P^{X}(dx)\times\delta_{m}(dy)\}$
and the OT problem becomes 
\[
\inf_{m\in\mathbb{R}}\int c(x,m)P^{X}(dx)=\inf_{m\in\mathbb{R}}\mathbb{E}[c(X,m)].
\]
For $c(x,y)=(x-y)^{2}$ one gets the squared $L^{2}$-Wasserstein
metric $W_{2}$ and the OT problem reads 
\[
W_{2}^{2}(P^{X},\delta_{m})=\inf_{m\in\mathbb{R}}\mathbb{E}[(X-m)^{2}]=Var(X),
\]
which is minimized for the mean $m=\mathbb{E}[X]$. For the $L^{1}$
distance $c(x,y)=\vert x-y\vert$ one gets the median, while for the
asymmetric cost $c(x,y)=(x-y)\alpha1_{x-y\geq0}+(y-x)(1-\alpha)1_{y-x>0}$
with $0<\alpha<1$ one gets the (left)-$\alpha$ quantile $m=q_{\alpha}(X)$,
i.e. the Value-at-Risk (see \cite{Koenker2005}. Finally, for $c(x,y)=y+\frac{(x-y)1_{x\geq y}}{1-\alpha}$
one gets the Conditional Value-at-Risk, namely the ES, see \cite{Rockafellar2002}.

In the magnitude-propensity approach investigated by \cite{Faugeras21},
the optimal transport approach scrutinizes traditional risk measure
limitations by reframing them as mass transportation from the original
risk distribution $P^{X}$ to a binary distribution $P^{Y}$, encapsulating
both risk magnitude and propensity. The corresponding risk measure
$(m,p)$ is derived by minimizing the Wasserstein $W_{2}-$distance
between $P^{X}$ and $P^{Y}$ within a constrained distribution set
${\cal P}_{0}:=\{P^{Y}=p\delta_{m}+(1-p)\delta_{0},p\in(0,1),m\in\mathbb{R}_{+}\}$.
This methodology provides a novel perspective on risk quantification,
integrating the multidimensional nature of risk into a cohesive framework,
where a loss of magnitude $m$ occurs with probability $p$,
and no loss (i.e. the loss amount equals zero) occurs with probability
$(1-p)$. It then offers a more nuanced approach to risk assessment,
potentially enhancing decision-making processes in financial and economic
contexts. If we denote with $Q_{X}$ (resp. $Q_{Y}$) the quantile
function associated with the distribution $P^{X}$ (resp. $P^{Y}$),
then the $L^{2}-$Wasserstein $W_{2}-$distance between $P^{X}$ and
$P^{Y}$ reads (see e.g. \cite{Rachev1998}): 
\[
W_{2}(P^{X},P^{Y})=\left(\int_{0}^{1}\left(Q_{X}(x)-Q_{Y}(x)\right)^{2}\,dx\right)^{1/2}
\]
where $Q_{X}(z):=\inf\{x:F_{X}(x)\geq z\},0<z<1$ and $F_{X}$ represents
the cumulative distribution function of the distribution $P^{X}$,
where $X$ is assumed to have finite variance. For the two-point distribution
$P^{Y}$, the quantile function reads $Q_{Y}(z)=m{\bf 1}_{1-p<z\leq1}$,
and the optimal quantities $(m,p)$ can be found explicitly
by direct optimization, see \cite{Faugeras21}. We omit the details
of their results as our work employs the quantization method, which
proves to be significantly more efficient in our context.  In Appendix \ref{appendixOT} we provide some additional details on the  algorithmic implementation of the Optimal Transport procedure that we adopted in our analysis.

\subsection{The Quantization Approach}

Quantization, rooted in engineering and signal processing, offers
an alternative paradigm for achieving analogous objectives. This approach
involves the optimal discretization of continuous risk distributions,
akin to analog-to-digital conversion and data compression methodologies.
The primary aim is to identify a finite set of points (codebook) that
minimizes the mean squared error or distortion between the original
risk distribution $P^{X}$ and its quantized counterpart $P^{Y}$.
This process effectively transforms complex, continuous risk landscapes
into more manageable, discrete representations.

Following \cite{graflush00} and \cite{LusPages23}, an $N$-vector quantizer on $(\mathbb{R}^{d},\vert\vert.\vert\vert)$
is a mapping $T:\mathbb{R}^{d}\rightarrow\{x_{1},...,x_{N}\}$ where
$\{x_{1},...,x_{N}\}$ is a codebook of size $N$, i.e. there is a
 partition (called Voronoi tessellation) $\{A_{i}\}_{1\leq i\leq N}$ with $A_{i}=\{x\in\mathbb{R}^{d}:T(x)=x_{i}\}$,
so that 
\[
T(x)=\sum_{i=0}^{N}x_{i}\mathbb{\mathbf{1}}_{A_{i}}(x).
\]
For a given $N\in\mathbb{N}$, an $N$-tuple of elementary quantizers
$(x_{1},\ldots,x_{N})$ is optimal if it minimizes over $(\mathbb{R}^{d})^{N}$
the quantization error: 
\begin{equation}
\|X-\hat{X}\|_{r}=\min_{(y_{1},\ldots,y_{N})\in(\mathbb{R}^{d})^{N}}\mathbb{E}\left[\min_{1\leq i\leq N}\|X-y_{i}\|^{r}\right]^{1/r}\label{gendistortion}
\end{equation}
induced by replacing $X$ by $\widehat{X}$. Then, instead of transmitting
the complete signal $X(\omega)$ itself, one first selects the closest
$x_{i}$ in the quantizer set and transmits its (binary or Gray coded)
label $i$. After reception, a proxy $\widehat{X}(\omega)$ of $X(\omega)$
is reconstructed using the code book correspondence $i\rightarrow x_{i}$.
Typically $r$ is fixed to be equal to 2, leading to a quadratic quantization
error. In this case, an $N$-optimal quantizer for a distribution
$P^{X}$ is a $N$-quantizer that minimizes the mean squared error
(also called the \textit{ distortion} function): 
\[
\inf_{T}\mathbb{E}[(X-T(X))^{2}].
\]
In $d$ dimensions, the minimal quantization error converges to zero
at a rate of $N^{-\frac{1}{d}}$ as $N\to\infty$, according to the
so-called Zador theorem. Several stochastic optimization procedures
based on simulation have been developed to compute these optimal quantizers.
For a comprehensive exposition of mathematical aspects of quantization, we refer to \cite{graflush00} and \cite{LusPages23}.
Remarkably, it can be shown (e.g. \cite{graflush00}) that an optimal
quantizer is a Monge map minimizing the Wasserstein metric $W_{2}(P^{X},P^{Y})$
between $P^{X}$ and $P^{Y}$, where $P^{Y}$ is a discrete measure
with $N$ points. In the case where $P^{Y}$ has two values, with
respect to the simplest quantization approach, we impose additional
constraints on the codebook points, with one point mass at zero and
another at a positive magnitude $m$, to indicate the presence of
loss. The quantization problem is then defined as the constrained
two-points quantizer with centers $\{x_{1},x_{2}\}:=\{0,m\}$, i.e.
as a mapping $T:\mathbb{R}^{+}\to\{0,m\}$ with 
\[
T(x)=\begin{cases}
m & x\geq a\\
0 & x<a,
\end{cases}
\]
where $a$ is a threshold to determine. Then, the optimal quantization
problem for $X\sim P^{X}$ with constrained knot at zero writes 
\[
\inf_{a,m\in\mathbb{R}^{+}}\mathbb{E}[(X-T(X))^{2}],
\]
and, as the boundary of the Voronoi cells separating the two quantizers
$x_{1},x_{2}$ is given by the center of the interval $[0,m]$, it
turns out that the Voronoi partition is given by the two regions $A_{1}=\{x\in\mathbb{R}_{+}:0\leq x\leq m/2\}$
and $A_{2}=\{x\in\mathbb{R}_{+}:x_{2}\geq m/2\}$. Thus, in the case
where $r=2,N=2,x_{1}=0,x_{2}=m$, \eqref{gendistortion} reads 
\begin{equation}
\|X-\hat{X}\|_{2}=\min_{m\in\mathbb{R}_{+}}\mathbb{E}\left[\min\{X^{2},(X-m)^{2}\}\right]^{1/2}
\end{equation}

For the two-point distribution for $P^{Y}$, \cite{Faugeras21} proved
the following result. 
\begin{thm}
(i) If $\mathbb{E}(X^{2})<\infty$ and the support of $P^{X}$ contains
at least two points, then there exists a magnitude-propensity pair
$(p_{X},m_{X})$ minimizing the distortion.

(ii) An optimal pair $(p_{X},m_{X})$ is characterized by solving for
$a$ the equation 
\[
2a=\mathbb{E}[X\vert X>a],a>0
\]
and then setting $m_{X}=2a,p_{X}=P^{X}(X>a)$. 
\end{thm}

Note that $m_{X}$ can be interpreted either as (twice) a $VaR_{\alpha}$
(resp. as an $ES_{\alpha}$) for a special value of the confidence
level $\alpha$: 
\[
m_{X}=CVaR_{p_{X}}(X)=2VaR_{1-p_{X}}(X).
\]
What is more, the confidence level $\alpha$ is no longer a subjective
choice dependent on the user. This is an improvement from a technical
perspective, possibly a drawback in the application playground, as
the prudential level is usually regulation-driven, i.e. assigned by
the authorities: typically, $\alpha=99\%$ in the market risk field,
$\alpha=99.9\%$ in the credit risk, finally $\alpha=99.5\%$ in the
\emph{Solvency} insurance regulation.

\subsection{Extension to the case of 3-Points Optimal Constrained Quantization}

In this subsection we are going to prove the main theoretical contribution
of the paper, namely the extension of the results of \cite{Faugeras21}
to the case of three points. The 3-Points distribution introduces
a sophisticated and comprehensive approach to risk quantification,
transcending the limitations of conventional methods. This distribution
represents risk through three distinct points. The first point is
the no loss case, $m_{0}=0$. The second point aims to designate moderate
risk $(m_{1},p_{1})$, where $m_{1}$ represents the magnitude of
potential moderate losses (corresponding to the most likely scenarios),
and $p_{1}$ represents the probability of occurrence for these losses.
The third point must capture extreme losses $(m_{2},p_{2})$, with
$m_{2}$ denoting their magnitude and $p_{2}$ indicating the probability
of occurrence for these severe events. As an example, the so-called
3-point risk analysis consists of summarizing the risk with 3 relevant
values, namely: the best case, the worst case, and the most likely
case. The three values are usually stimated (assessed) by submitting a standardized
questionnaire to a panel of experts. For an extended review of these
qualitative risk assessment techniques, see e.g., \cite{aven2016risk}.
The quantitative approach of our work differs from this kind of methodology
for its more rigorous and objective framework. The 3-Points distribution
with a specific constraint introduces an additional criterion that
influences the optimal selection of the discrete distribution parameters.
This modification aims to avoid that the pure mathematical optimal
solution does not highlight high magnitude losses. Furthermore, the
stability of a risk measure over time is a general requirement of
any risk measure in the banking regulation.

Following \cite{graflush00}, a $3$-vector quantizer on $\mathbb{R},\vert\vert.\vert\vert)$
is a mapping $T:\mathbb{R}\rightarrow\{x_{0},x_{1},x_{2}\}$, where
we relabeled the codebook ($\{x_{1},x_{2},x_{3}\}\rightarrow\{x_{0},x_{1},x_{2}\}$)
for notational convenience. More specifically, one can refine a measure
of risk into a classification between a ``moderate'' and a ``large''
loss, by using a three points discrete measure, $P^{Y}=(1-p_{1}-p_{2})\delta_{0}+p_{1}\delta_{m_{1}}+p_{2}\delta_{m_{2}}$,
with $0<m_{1}<m_{2}$ and where $x_{0}=m_{0}=0$. Utilizing a three-point
discrete measure enables the encoding and quantification of both moderate
and large losses on the magnitude and propensity scales. This is achieved
through the pairs $(m_{1};p_{1})$ and $(m_{2};p_{2})$, respectively.
One thus introduces the constrained three-point quantizers with centers
$\{0,x_{1},x_{2}\}:=\{0,m_{1},m_{2}\}$, i.e. as a mapping $T:[0,\infty)\rightarrow\{0,m_{1},m_{2}\}$
with $T(x)=0{\bf 1}_{x\leq a_{1}}+m_{1}{\bf 1}_{a_{1}<x\leq a_{2}}+m_{2}{\bf 1}_{x>a_{2}}$,
where $0\leq a_{1}<a_{2}$ are thresholds to be determined. Then, the
optimal quantization problem with constrained knot at zero for a random
variable $X$ with $\mathbb{E}[X^{2}]<\infty$ (and whose support
contains at least three points) writes 
\begin{align}
\inf_{(a_{1},m_{1},a_{2},m_{2})\in[0,+\infty)^{4}}\mathbb{E}[(X-T(X))^{2}].
\end{align}
Using standard arguments, see e.g.  \cite{graflush00} and \cite{LusPages23}, one already knows that $a_{1}=m_{1}/2$ and $a_{2}=(m_{1}+m_{2})/2$,
i.e. that the Voronoi regions write 
\begin{align}
A_{0} & =\{x:0\leq x\leq m_{1}/2\},\label{A0}\\
A_{1} & =\{x:m_{1}/2\leq x\leq(m_{1}+m_{2})/2\},\label{A1}\\
A_{2} & =\{x:x\geq(m_{1}+m_{2})/2\}.\label{A2}
\end{align}
As a consequence, the distortion/objective function writes as a sole
function of the magnitudes $m_{1},m_{2}$ as 
\begin{align}
D(m_{1},m_{2}) & =\mathbb{E}[X^{2}\wedge(X-m_{1})^{2}\wedge(X-m_{2})^{2}]=\mathbb{E}[X^{2}\wedge(X-m_{2})^{2}]{\bf 1}_{m_{2}\geq m_{1}},\label{distortion}
\end{align}
since $m_{2}>m_{1}$. Introduce now for $x=(x_{1},x_{2})\in\mathbb{R}_{+}^{2}$
\begin{align*}
H(x_{1},x_{2}) & =h_{0}(x_{1},x_{2})\wedge h_{1}(x_{1},x_{2})\wedge h_{2}(x_{1},x_{2}):=X^{2}\wedge(X-x_{1})^{2}\wedge(X-x_{2})^{2}.
\end{align*}

\begin{figure}
\centering \includegraphics[scale=0.6]{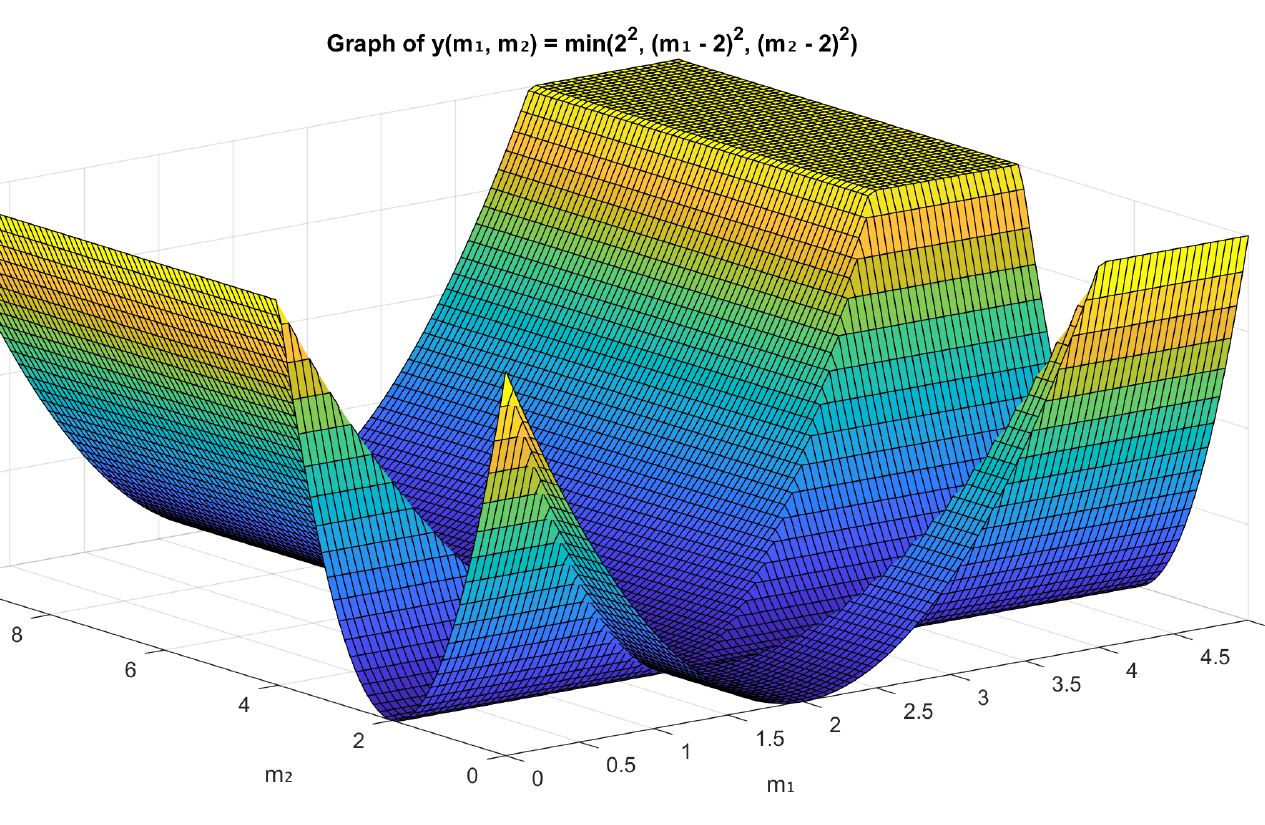} \caption[]{Graphic of the function $H(m_{1},m_{2})=2^{2}\wedge(m_{1}-2)^{2}\wedge(m_{2}-2)^{2}$.}
\label{minima}
\end{figure}

Note that $h_{0},h_{1},h_{2}$ are differentiable functions, so that
the function $H$ is (at least) piecewise differentiable, and it has
directional derivatives everywhere, as shown in Figure \ref{minima}
where we display the function $H$ for $X=2$. 

As shown in \cite{Faugeras21}, a convenient tool in the setting of
optimization of piecewise smooth functions is the concept of Bouligand
derivative (B-derivative), see e.g. \cite{Scholtes12}. By dropping the
requirement of linearity of the differential, it represents a first-order
approximation and allows to have a single-valued notion of differential.

\defn A function $f:\mathbb{R}^{2}\rightarrow\mathbb{R}$ is B-differentiable
at $\tilde{x}\in\mathbb{R}^{2}$ if there exists a positive homogeneous
function $\nabla^{B}f(\tilde{x}):\mathbb{R}^{2}\rightarrow\mathbb{R}$
s.t. 
\begin{align}
f(\tilde{x}+v) & =f(\tilde{x})+\nabla^{B}f(\tilde{x})(v)+o(\vert\vert v\vert\vert),\ \ \forall v\in\mathbb{R}^{2}.
\end{align}

We have the following lemma on the B-differentiability of the function
$H$.

\lem Let $H(x)=\min(h_{0}(x);h_{1}(x);h_{2}(x))$, where $h_{0},h_{1},h_{2}:\mathbb{R}^{2}\rightarrow\mathbb{R}$
are differentiable functions. Then $H$ is B-differentiable, and its B-differential is given as follows.
\begin{enumerate}
\item For $i,j,k$ distinct indexes in $\{0,1,2\}$, for $x$ such that $h_{i}(x)<\min(h_{j}(x);h_{k}(x))$ 
\begin{align*}
\nabla^{B}H(x)(v) & =\nabla h_{i}(x)v=\frac{\partial h_{i}}{\partial x_{1}}(x)v_{1}+\frac{\partial h_{i}}{\partial x_{2}}(x)v_{2}.
\end{align*}
\item For $x$ such that $h_{i}(x)=h_{j}(x)<h_{k}(x)$  
\begin{align*}
\nabla^{B}H(x)(v) & =\min(\nabla h_{i}(x)v;\nabla h_{j}(x)v).
\end{align*}
\item For $x$ such that $h_{0}(x)=h_{1}(x)=h_{2}(x)$  
\begin{align*}
\nabla^{B}H(x)(v) & =\min(\nabla h_{0}(x)v;\nabla h_{1}(x)v;\nabla h_{2}(x)v).
\end{align*}
\end{enumerate}
\label{lemmaH}

\proof See  Appendix \ref{appendixProofs}.

Before proving the B-differentiability of the distortion function
$D$, we recall that a point $x$ is critical (that is, a potential minimizing point)  for $D$ if $ \nabla^{B}D(x)(v)\geq0\ \forall v\in\mathbb{R}^{2}$.

\thm The distortion function $D$ defined in \eqref{distortion}
for a random variable $X\in L^{2}$ is B-differentiable on $x\in\mathbb{R}_{+}^{2}$,
with B-derivative given by 
\begin{align}
\nabla^{B}D(x_{1},x_{2})(v) & =\mathbb{E}[2(x_{2}-X)v_{2}{\bf 1}_{x_{1}+x_{2}<2X}]\nonumber \\
 & +2\mathbb{E}[\min((x_{1}-X)v_{1}{\bf 1}_{x_{1}<2X}{\bf 1}_{x_{1}+x_{2}>2X}]\nonumber \\
 & +2\mathbb{E}[\min(0;2Xv_{1}){\bf 1}_{x_{1}=2X}].\label{BD}
\end{align}
Moreover, the points $(0,0),(0,2X),(2X,0),(2X,2X),(0,x_{2})$ (with
$x_{2}>2X$) and $(x_{1},0)$ (with $x_{1}>2X$) are not critical
for the distortion function, since $\nabla^{B}D(x)(v)<0$ for some
$v\in\mathbb{R}^{2}$.\label{teoremaD}

\proof See Appendix \ref{appendixProofs}.


We have now all the ingredient to characterize the critical points,
namely we can find the quantizers of the 3-points distribution by
solving a system of equations. From Equation \eqref{BD}, by taking
separatly $v_{1}=0,v_{2}=0$ we get 
\begin{align*}
P\left(X=\frac{x_{1}+x_{2}}{2}\right)=P\left(X=\frac{x_{1}}{2}\right)=0,
\end{align*}
and we find the following system 
\begin{align*}
\mathbb{E}[(x_{1}-X){\bf 1}_{\frac{x_{1}}{2}<X<\frac{x_{1}+x_{2}}{2}}] & =0,\\
\mathbb{E}[(x_{2}-X){\bf 1}_{X>\frac{x_{1}+x_{2}}{2}}] & =0,
\end{align*}
which leads to the optimal quantizers: 
\begin{align}
m_{1} & =\mathbb{E}\left[X\vert\frac{m_{1}}{2}<X<\frac{m_{1}+m_{2}}{2}\right],\label{m1}\\
m_{2} & =\mathbb{E}\left[X\vert X>\frac{m_{1}+m_{2}}{2}\right].\label{m2}
\end{align}
Eventually, once the system is solved using a fixed-point technique,
the probabilities $p_{1},p_{2}$ associated with the quantizers are
given by integrating the probability of $X$ on the Voronoi regions
given by \eqref{A1}--\eqref{A2}. Finally, the probability mass
of the initial (constrained) quantizer $m_{0}=0$ is given by $(1-p_{1}-p_{2})$.

\begin{Remark}
The points $m_1$ and $m_2$
 optimally represent the underlying loss distribution, independently of any $a\ priori$ thresholds such as those described  by $VaR$ or $ES$ associated with some confidence level. Consequently, there is no guaranty that $m_2$, intended to represent an extreme loss, can be directly compared to $VaR$ or $ES$, nor that the associated probability 
$p_2$ has any direct relationship to the confidence level $\alpha$. Therefore, it is not surprising that, as we show in our numerical experiments, the value of $m_2$ can differ from  $VaR$. In other words, comparing $m_2$ with $VaR$ is not a meaningful exercise, as we only have  one $m_2$ while we can compute many $VaR$ according to  the relevant confidence levels adopted by the financial regulation used in practice (0.95, 0.99, 0.995, 0.999, etc.).
Nevertheless, it is advisable to introduce a constraint to prevent the purely mathematical optimum from neglecting high-magnitude losses and to ensure the temporal stability of the risk measure, which is a requirement in banking regulation. In the context of market risk, where $m_2$
 typically turns out to be slightly lower than $VaR$, it seems reasonable to impose an additional restriction on $m_2$, ensuring that it reaches a value at least equal to $VaR$. This preserves its interpretation as an indicator of extreme risk while maintaining compliance with regulatory standards, which are expressed in terms of $VaR$.
From an analytical standpoint, the introduction of such a constraint increases the complexity of the optimization procedure.  However, from a numerical perspective, the procedure remains highly efficient (see the appendix for further details on the optimization procedures), and the resulting performance is fully satisfactory.\end{Remark}

\section{Case Study based on Real Datasets}

\subsection{Practical Risk Management}
In this section we try to exploit the features of the magnitude-propensity approach in relevant different fields.
To achieve significant insights, some concepts about practical risk management are needed.
First, the \textit{risk category} must be well defined.
 In a high level classification, we distinguish market risk vs. credit risk. Market risk is given by the PnLs (profits and losses) uncertainty due to the price dynamics of the assets in the portfolio. Financial portfolios of large banks typically are not concentrated; even if we may observe extreme events due to volatility peaks, the PnLs typical distribution is quite smooth.
 Credit risk is related to the losses implied by any default in the portfolio, where once the default occurs, the recovery rate drives the amount that one can get back. In this case, we have also concentrated portfolio, and the binary nature of the outcome (no default, default) implies a loss distribution that sometimes shows some peaks in the extreme losses region.
 Due to high quality, high frequency data and finally a long history of the risk management in the financial area, the regulation typically asks for more conservative confidence level in the credit risk (e.g. 0,999), as we have low frequency data, low data coverage (e.g. many issuers in the portfolio might be unrated, with missing default probability), and the default correlations are not strict sense observable.
 The above issues are summarized in the model risk definition, i.e. the risk arising from misspecifications in the model itself or in its parameters.
 The boundary between market and credit risk is quite flexible. A relevant example is given by the \textit{migration risk}, i.e. the risk of losses due to downgrade (rating worsening) of one or more issuers in the portfolio.
 Furthermore, the well known \textit{spread risk}, i.e. the risk of losses due to the increase of the spread level (e.g. in the government bonds in portfolios) is clearly assigned to the market risk discipline in the banking context, see \cite{bcbs2019fundamental}, while it can be allocated to market or credit risk filed according to the Solvency regulation for the insurance companies, see \cite{Solvency}.
 See \cite{Frey} for a comprehensive survey of the main modeling methodologies related to market and credit risk.
Taking into account the above practical and regulatory concepts, we decided to exploit the capabilities of the magnitude-propensity 3-points proposal in three different application contexts, trying to cover a relevant area of the broad market and credit risk fields, as summarized in Table \ref{tab:risk_cases_overview}.

\begin{table}[h!]
\centering
\small
\renewcommand{\arraystretch}{1.15}
\setlength{\tabcolsep}{3pt} 

\resizebox{0.9\textwidth}{!}{ 
\begin{tabular}{
    c
    p{40mm}
    p{25mm}
    p{20mm}
    p{40mm}
    p{20mm}
    p{12mm}
}
\toprule
\textbf{ID Case} &
\textbf{Description} &
\textbf{Risk Category} &
\textbf{Regulation} &
\textbf{PnLs Model} &
\textbf{Confidence level} &
\textbf{Horizon} \\
\midrule
1 & Market Risk Trading book & Market Risk & Basel & Historical non parametric & 99\% & 1D \\
2 & Credit Risk Trading Book & Credit Risk & Basel & Monte Carlo parametric &  99{.}9\% & 1Y \\
3 & Market Risk Insurance & Market Risk & Solvency & Monte Carlo parametric  & 99{.}5\% & 1Y \\
\bottomrule
\end{tabular}
} 

\caption{Overview of VaR calculation according to regulatory risk cases: model type, regulation, model calculation, confidence level and horizon.}
\label{tab:risk_cases_overview}
\end{table}

\subsection{Applications Overview and Preliminary Concepts}

Financial Institutions commonly adopt two main approaches to measure
the risk of a financial position, which they select based on their
specific purposes and regulatory requirements: \textit{sensitivity} measures
and risk measures based on the profit and loss (PnLs) distribution. Both these approaches are prescribed by the regulations and adopted by the risk management departments.

To have a better understanding, let us denote the value of a generic
portfolio at the evaluation time $t_0$ as $V_{t_0}$. Following standard
risk-management practice, $V_{t_0}$ is modeled as a function of  time
$t$ (the index $t_0$ represents the current time, so that e.g. $t_0-1$ indicates the previous time and so on) and a $d-$dimensional random vector ${\bf Z}_{t}=(Z_{t}^{1},Z_{t}^{2},\cdots,Z_{t}^{d})$
of risk factors, i.e., underlying variables or drivers like e.g. interest
rates, stock prices, implied volatility, or exchange rates. Hence,
\begin{equation}
V_{t_0}=f(t_0,{\bf Z}),
\end{equation}
with $f:\mathbb{R}\times\mathbb{R}^{d}\rightarrow\mathbb{R}$ and where ${\bf Z}$ informally represents the heterogeneous set of risk factors (scalar, array, surface, cubes etc.) recorded at past observation times, depending on specific payoffs of the instruments of the portfolio.

For the sake of simplicity, we omitted in the expression the various parameters on which depend the stochastic processes that describe the risk factors $\textbf{Z}_t$ dynamics. 
\textit{Sensitivity} based risk measures rely on each risk factor impact, which
can be defined as the change in value of an instrument (position) given a small, predetermined (hypothetical) movement in a risk factor that affects
the instrument's value: from a mathematical perspective, sensitivities
are computed as partial derivatives of the function $f$ with respect
to $Z$. By defining $\Delta{\bf Z}= {\bf Z}_t-{\bf Z}_{t-1}$ the vector of observed shocks applied to the risk factors, this approach enables the estimation, assuming linearity between $V$
and ${\bf Z}$, of the portfolio response to the new market conditions  by multiplying $\Delta{\bf Z}$
with the corresponding factor sensitivities, namely $\Delta V\cong\nabla f\cdot{\bf \Delta Z}$.
Technically speaking, it is just a differential of the value function
$f$ evaluated for some small  increment $\Delta{\bf Z}$. Of course, the second order impact could be added to improve the accuracy of the approximation. The most popular second order approximations are referred according to some naming conventions, such as the \textit{gamma} effect, i.e. the second order derivative of an option the the underlying price, and the \textit{convexity}, namely the second order impact of the interest rate level change on a bond or interest derivative (e.g swaps, caps and floor options).
Such risk measures, though valuable in providing information about
the robustness of a portfolio value to specific events, have limitations
when making capital-adequacy decisions: they do not deal with the
dependency properties of the risk drivers, hence they can not create
a picture of the overall riskiness of the portfolio of a financial
institution, and they do not have any information about the likelihood
of the approximated \emph{PnLs}, see \cite{mcneil2005quantitative}. Furthermore, sensitivity approach is a typical \textit{what-if} approach, where there is some subjectivity in defining the level of the extreme market shock to be applied.
For this reason, the so called \textit{probabilistic} measures, such as $VaR$ and $ES$, are prominent in the financial regulation, to ensure that banks and the insurance companies have enough own capital to face the potential losses under "extreme yet plausible" scenarios, a statement adopted very often  by the financial authorities to define the scope of the stress test exercise.   The following subsections give more details on the context of the three applications of the methodology.

\subsubsection{Market Risk with Historical Simulation}
In the introduction, we recalled that the historical simulation is
the most popular approach in the large banks to evaluate the financial
risk, by \emph{VaR}, \emph{ES} or any other measure. Let us give a
brief formal explanation, mainly of the historical simulation
through the \emph{full evaluation} methodology. We can define the daily portfolio
\textit{PnLs} as the change in value of the portfolio, driven by the
series of risk factor changes ${\bf Y}_{t}$ where ${\bf Y}_{t}:={\bf Z}_{t}-{\bf Z}_{t-1}=\Delta {\bf Z}$
or alternatively ${\bf Y}_{t}:=\ln\left({\bf Z}_{t}/{\bf Z}_{t-1}\right)$.
The \textit{additive} (or absolute) vs \textit{multiplicative} (or percent) definition of 
a shock is calculated depending on the asset class of the risk factor: usually,
banks adopt the additive convention for interest rates and spreads, while
the multiplicative convention for returns on equities, foreign exchange rates and funds. For a useful survey, see e.g. \cite{Hudson2010}. 
For each $t=t_0-1,\cdots,t_0-S$ in our array of historical scenarios, the
portfolio value is calculated by summing up the contributions over all proper pricing functions $f_m$ (related to the instrument $m=1,\cdots, M$) that select just the required risk factors, typically a small   subset of the whole market
data ${\bf Z}$.

If we adopt the \textit{full evaluation} approach, the
portfolio's \textit{PnL} at a given calculation time ($t_0$) and for any observed scenario
$[t,t_0]$ with $t=t_0-1,\cdots,t_0-S$, is given by: 
\begin{equation}
PnL_{[t,t_0]}=\sum_{m}f_{m}(t_0,{\bf Z}+{\bf Y}_t)-f_{m}(t_0,{\bf Z}).\label{PnL}
\end{equation}

The value (or \emph{PnLs}) of the portfolio is then the sum of a huge
variety of pricing functions with heterogenous risk factor inputs.

\begin{Remark}
The above stylized expression uses the additive convention for the shock. In the multiplicative case the part $f_{m}(t_0,{\bf Z}+{\bf Y}_{t})$
is replaced by $f_{m}(t_0,{\bf Z}\cdot\exp({\bf Y}_{t}))$.
\end{Remark}

\begin{Remark}
In the above expression we refer to absolute (eur, dollar, etc.) \textit{PnLs}, i.e. the standard convention for bank portfolios. In the asset management field it is common to switch to relative (percent) \textit{PnLs}, by a scaling factor, i.e. the current value $V_{t_0}$ of the portfolio.
\end{Remark}

\begin{Remark}
If  \(PnL\) is calculated using the sensitivities approach,  the function $f_m(. )$ is replaced by the gradient vector, and we have $PnL\cong\nabla f_m\cdot{\bf \Delta Z}$, possibly enriched with the second order term based on the Hessian matrix.
\end{Remark}
\begin{Remark}
One could replace the term $f_{m}(t_0,{\bf Z}+{\bf Y}_{t})$ by the more general expression $f_{m}(t_0+1,{\bf Z}+{\bf Y}_{t})$ 
to properly take into account the ``ageing effect'', meaning that  if the
market shocks are supposed to prevail in a given time horizon $h$ (  $h=1$ for simplicity in this case), then the hypothetical \emph{PnLs}
must take into account such time shift in order  to shorten the time to maturity of the instruments, such as bonds or derivatives. Despite this seems an obvious
concept, very often the ageing is not actually considered, as it would
require that all the pricing libraries of the bank are able to price
in the future, i.e. not only taking into account the time decay (such
as the decreasing maturity for a bond) but also all the other possible events,
such as dividend payments and so on. In most cases, a  risk factor shock is applied, but not the time shift, so applying the \emph{instantaneous shock} assumption. For the insurance sector, see the survey in \cite{EIOPA}, where it is outlined that only 6
out of 20 insurance companies manage the ageing effect, while the
remaining ones  apply the instantaneous shock model. 
\end{Remark}

In other words, \textit{PnLs} distribution consists in the range of
possible values $X$ (potential profits and losses) that a portfolio
may experience over a specific time horizon. 
The historical simulation approach is data driven, as no parametric distribution family is assumed for the returns. Furthermore, we do not need any correlation parameters or dependence modeling as it is assumed that the dependencies among the risk factors are divided into small pieces inside each scenario $(t)$.
Recalling that in measuring
portfolio's risk the fundamental aspect is estimating the cumulative
density function of the \textit{PnLs}, namely $F(x)=\mathbb{P}(X\leq x)$,
or functionals of it, see \cite{mcneil2005quantitative} and that Value at Risk is essentially a
quantile (typically $95\%,99\%,99.9\%$) of the above mentioned \textit{PnLs}
distribution, the final step consists in estimating  a suited percentile  to be applied to the vector
of \textit{PnLs}. For the number $S$ of scenarios,  banks usually adopt $S=250$ or $S=500$, i.e. they
collect about 1 or 2 years of daily changes in all the risk factors to which the portfolio is sensitive. The regulatory time horizon $h$ differs
from the daily popular horizon, e.g. $h=10$ days in the current Basel
regulation. Collecting a sample of historical non-overlapping risk
factors returns $\left( {\bf Y}_{t,h}={\bf Z}_{t+h}-{\bf {\bf Z}}_{t}\right)$
could be quite difficult. As an example, to achieve $S=250$ non overlapping scenarios with $h=10$ days, we need $h \cdot\ S=2500$ observations, i.e. around 10 years of full time series for all the risk factors.
Hence, the authorities generally accept the
\emph{square root} rule, namely $VaR\left(\alpha,h\right)\equiv VaR\left(\alpha,1\right)\cdot\sqrt{h}$.

In daily practice, \textit{VaR} can be estimated in many ways: from the
basic \textit{empirical} percentile to some more robust estimators, such as
$L-$estimators and Harrel-Davis estimator, that smooth the estimation
by averaging the \textit{PnLs} in a neighborhood of the empirical
quantile, see\cite{Harre1982}. In this field a comprehensive reference is given by \cite{David2004}. These more advanced estimators of course may be used also in the other simulation contexts, such the Monte Carlo simulation.
The historical simulation approach is very popular in the banking industry
due to some key reasons. First, it is very intuitive, as the
empirical past (recent) distribution of the risk factors returns is
accepted (assumed) to be the best estimation of the unknown exact distribution.
No functional subjective assumption is made about the distribution
shape. Being purely data-driven, it does not rely on any parameters
(volatility, correlations, etc). In real world, portfolios may
contain many thousands of risk factors: estimating the related parameters
and updating periodically the estimates is a very challenging 
task. For the sake of simplicity, in the following applications we
will refer to the basic empirical quantile (e.g. the $5th$ worst
result in an array of $S=500$ $PnL$).

\subsubsection{ Credit Risk in the Trading Book: the Default Risk Charge by Montecarlo approach}

The Default Risk Charge (\emph{DRC}) is a regulatory measure designed
to capture default risk within the trading portfolio, as required
by Basel standards, particularly within the framework of the Fundamental
Review of the Trading Book (FRTB) outlined in the \cite{bcbs2019fundamental}
document. This model is designed to quantify the risk of loss resulting
from the failure of a counterparty or issuer of financial instruments,
including equity, bond, and derivative exposures. The Default \emph{DRC}
is specified in Chapter 7, and in the updated Basel Framework,  in Paragraphs MAR 33.18- 33.38. These documents establish the criteria
for calculating default risk, specifying that
\begin{itemize}
\item It must be calculated over a one-year horizon. 
\item It must reflect a 99.9\% confidence level.
\item It must include all exposures sensitive to default risk within the
trading portfolio, excluding those specifically defined as \textit{non-material}
risks.
\end{itemize}
In the practice, the default of each issuer is modeled by a Merton
type model. The \textit{credit worthiness} dynamics of the obligor $n\ (n=1,...,N)$
is defined by the model 
\begin{equation}
\Delta X_{n}=\sum_{k=1}^K\beta_{k,n}\cdot W_{k}+\sigma\cdot\varepsilon_{_{n}}
\end{equation}
where the coefficients $\beta_{k,n}$ are the \emph{factor loadings}
and describe the systematic risk, driven by a set of risk factors
$k=1,...,K$, the term $\varepsilon_{n}$ is the specific (obligor) factor,
with $W_{k}$ correlated standardized Normal random variables, with $E\left[W_{k}\varepsilon_{n}\right]=0\ \forall n,k$.

In this standardized framework, the default happens if the credit
worthless is below a given threshold, $D_{n}=\left\{ X_{n}\leq\Phi_{n}^{-1}\left(PD_{n}\right)\right\} $,
where $\Phi_{n}$ and $PD_{n}$ indicate the cumulative distribution
of $\Delta X_{n}$ and the default probability of the $n-th$ issuer
in the portfolio. The factor loading coefficients are estimated by
a statistical regression step, by combining the time series of
the list of risk factors with the issuer equity prices or spreads.
In large banks, we generally have hundreds of issuers ($N$) and some
dozens of risk factors ($K$). The distribution of the portfolio loss is given by \ref{DRC}, that we recall here:
\begin{equation*}
Loss_{DRC}=\sum_{n=1}^NEAD_{n}\cdot\mathbb{\mathbf{1}}_{ D_{n} }\cdot LGD_{n}.
\end{equation*}
This quantity 
may not be available in closed form, thus most banks develop
Montecarlo simulation algorithms. Indeed, also the Basel regulation
explicitly refers to the simulation approach. Due to the extremely high confidence
level $\alpha=99.9\%$, to achieve a sufficient accuracy the number
 of simulation scenarios must be very high, usually in the range $\left[100k,1Mln\right]$.
At the end of the simulation cycle, the quantile is estimated (empirically or using some smoother estimator, in line with the usual approach adopted   for computing the \emph{VaR} by historical simulation).

With respect to the work by \cite{Faugeras21}, where the authors deal
with the losses coming from the claims of the clients (i.e.,  the domain of
the distribution lies in one-side in the real axis), we recall that for the \emph{VaR}
of the financial portfolio, a bank can experience both profits and
losses, while for the \emph{DRC} we just have losses, so making the
context identical to that of the insurance sector. 

\subsubsection{ Market Risk in the Insurance field: Monte Carlo approach}\label{SectionInsurance}

While in the banking sector the liabilities side of the balance sheet typically mirrors the asset side—comprising the same types of instruments such as bonds, equities, derivatives, and loans with opposite signs—in the insurance business the situation is structurally different. The asset side, usually consisting of bonds and investment funds, is primarily designed to hedge the liabilities, namely the claims arising from policyholders’ underwriting activity.

Since the products sold to clients span very heterogeneous categories (life, longevity, accident, financial, and others) and often include numerous complex clauses linked to a wide range of contingent events, it becomes extremely challenging to construct a fully granular model incorporating all the specific inputs. For this reason, most insurance companies adopting a probabilistic approach to risk modeling rely instead on a compact set of risk factors, which are expected to adequately capture the distribution of profits and losses across the entire balance sheet.

To do that, the insurances use an \textit{economic scenario generator} (ESG) that adopts the Monte Carlo simulation to project thousands of paths of the selected risk factors. For each path, all the positions of the insurance are evaluated to obtain the profile of profits and losses. Finally, the desired risk measures (percentiles, expected shortfall, etc) are estimated by analyzing the simulated results.
For a recent extended review see \cite{EIOPAESG}.
Briefly, the Monte Carlo approach is not a choice in the insurance sector, but it represents a mandatory tool, due to the high number of risk factors and to the complexity of the products managed in insurance.
Of course, also in the practical applications several improvements have been implemented to reduce the weaknesses of the basic Monte Carlo approach, such as the simulation error. 
The improvements are related to both the simulation step (by low discrepancy algorithm, quasi Monte Carlo methods, etc) and the estimation step, by smoothing the straight empirical estimator with more robust estimators, e,g, the Harrel-Davis estimator. 
In our application, we collected samples of 100K or 200K simulations related to the market risk of a major European insurance company and compared the 3-points results, mainly $m_1$ and $m_2$, with respect to the 0.995 1 year Solvency $VaR$.

\subsection{The Dataset}

The data on which we apply the approach presented in this paper includes
1 year of daily \textit{PnLs} time series of a representative portfolio
of a large European bank, encompassing both its banking and trading
books. As explained in the previous section, the \textit{PnLs} have
been obtained by full revaluation of the positions composing the portfolio
with historical simulations of the risk factors. The time window includes
254 business dates (July 3rd 2023- July 1st 2024).

The portfolio consists of more that 100k positions, spanning from
bonds to equities to derivatives. A couple of business dates where
some outliers were detected (due to failures in the bank software
systems) have been removed. It is worth to note that the process to
get the portfolio \textit{VaR} is typically a bottom-up process. In
other words, Formula \ref{PnL} is just a synthetic definition, as
it is practically calculated as 
\begin{equation}
PnL_{[t,t_0+h]}=\sum_{m=1}^{M}f_{m}(t_0+h,{\bf Z}+{\bf Y}_{t+h},{\bf P}_{n}),
\end{equation}
being $n$ the index of each position in the portfolio and ${\bf P}_{n}$
the information (maturity, coupon, currency, etc) related to any position.
The \emph{VaR} in historical simulation (Case 1) is then usually reported at
a portfolio level, but it can be easily broken down for any purpose
at any more granular level, according to the different analysis that
are required.

As concerns the DRC (Case 2), we have for the same bank a set of some end-of-month calculation, referred to March 2024, June 2024, December 2024
and to February 2025. The outcomes are based on $200k$ simulations.
The portfolio consists of many thousands positions, belonging to $O\left(10^{3}\right)$
issuers chat could default. Finally, for the insurance portfolio (Case 3), we analyzed an array of $100k$ PnL generated by a multi dimensional Gaussian copula based on about $150$ risk factors. 

\bigskip

To test the methodology and the numerical procedures in a more comprehensive context, we exploit the methods for the three fields defined in the previous section.

\begin{itemize}
   \item Case 1: Market Risk in the Banking sector. In this case the methodology for the PnL distribution is based on the historical simulation with 250 scenarios. Regulatory VaR (Basel regulation): 1 day, 99\% confidence level.
Risk sources: equity, interest rates, forex, spread, commodities.
   \item Case 2: Credit Risk in the Banking sector. Then we refer to binary events driven by PD (Default Probability), that determine the losses by the LGD (loss given default) parameter applied to the defaulted position. LGD could be stochastic, e.g. a Beta random variable in the $[0,100\%]$ range. Methodology for the PnL distribution: Montecarlo simulation, multivariate gaussian copula, Merton-type default model. Regulatory VaR (Basel regulation): 1 year, 99.9\% confidence level. Risk source: default. 
   \item Case 3: Market Risk in the Insurance sector. Here the methodology for the PnL distribution is based on Montecarlo simulation, multivariate gaussian copula, and Merton-type default model. Regulatory VaR (Solvency regulation): 1 year, 99.5\% confidence level. Risk sources: equity, interest rates, forex, spread, credit migration.
\end{itemize}

\subsection{Results}

In this subsection we will show that overall, the results demonstrate that our methodology based on the magnitude propensity remains theoretically sound, numerically stable, and flexible enough to account for diverse risk profiles, while preserving interpretability across market, credit, and insurance domains. 

\subsubsection{Case 1: Market Risk in the Banking Sector}

\begin{figure}[h!]
\centering
\includegraphics[width=0.8\textwidth]{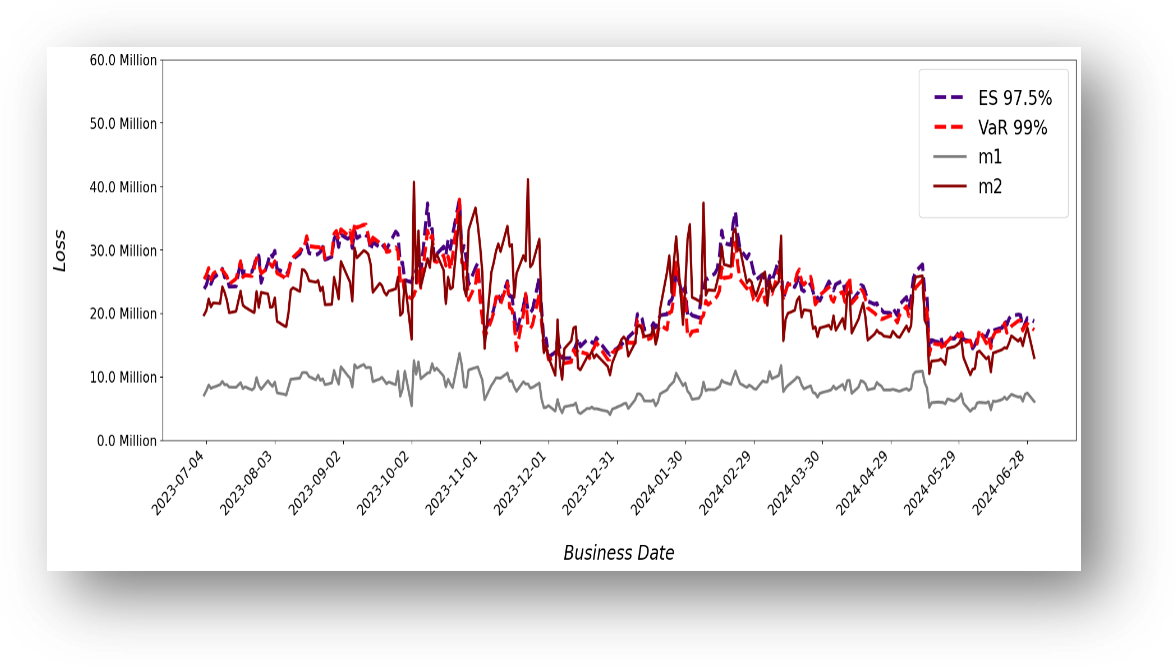} 
\caption{Time series plot of VaR 99\%, ES 97.5\%, $m_1$ and $m_2$ (without constraints on $m_2$). }
\label{fig:market risk unconstrained}
\end{figure}

In the initial stage of analysis, we consider the analytical solution obtained by quantization using the fixed-point method, allowing the quantities $m_1$ and $m_2$  to vary freely, i.e. subject only to the natural constraint 
$0<m_1<m_2<worst\ case$. Figure \ref{fig:market risk unconstrained} illustrates the daily evolution of $m_1$ and $m_2$, and compares them with the conventional risk measures—$VaR$ at the 99\% confidence level and ES at the 
97.5\% level, the latter chosen to ensure comparability with the 99\% VaR under a Gaussian framework. The unconstrained three-point distribution provides a clear and interpretable structure for quantifying risk, yielding results of a conservative nature that capture both moderate and extreme losses while avoiding excessive overestimation.

The interpretation of $m_1$ as a measure of moderate risk appears straightforward; however, the parameter $m_2$ 
 often exhibits overly conservative behavior, frequently taking values below the corresponding VaR. This outcome is expected, as the VaR constraint becomes more binding in market environments characterized by moderate tail risk and smoother loss distributions. Consequently, the resulting estimates tend to be slightly conservative, consistent with the regulatory framework that defines VaR at the 99\% confidence level over a one-day horizon. To maintain coherence with this framework—where VaR is associated with extreme risk levels—we subsequently impose a constraint requiring $m_2$ 
 to exceed the VaR threshold.

\begin{figure}[h!]
\centering
\includegraphics[width=0.8\textwidth]{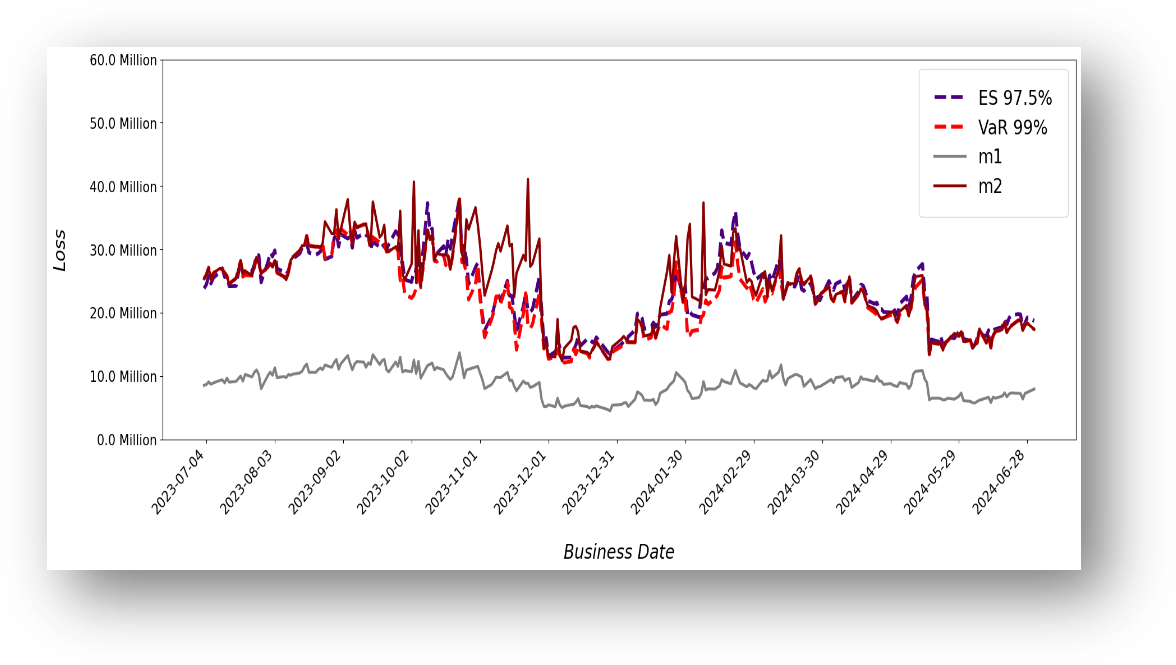} 
\caption{Time series plot of VaR 99\%, ES 97.5\%, $m_1$ and $m_2$ (with constraint $m_2>VaR$).}
\label{fig:market risk constrained}
\end{figure}

If we add a constraint on the $m_2$, the methodology provides the most satisfactory results, see Figure \ref{fig:market risk constrained}.

 The
magnitude parameters $m_1$ and $m_2$ are intuitive, capturing both moderate and extreme
losses, with stable dynamics over time relative to VaR and ES. Meanwhile, $p_1$ and $p_2$
remain smooth, with $p_2$ constrained by $m_2 > VaR$ and $p_1$ free to vary within the $(0,1)$
range.

 As concerns the volatility of $m_2$, this naturally follows from the variability of the VaR computed by the historical simulation method. Indeed, when the scenarios contributing to the empirical percentile fall outside the rolling window used for VaR, jumps  may occur independently of the portfolio composition. This phenomenon, known as the \textit{reshuffling effect}, could be mitigated or controlled through smoothing techniques, in analogy with those employed in filtered historical simulation, see e.g. \cite{Barone1999}.

Furthermore, Table \ref{tab:var_comparison_market_risk_compact} shows that employing different optimization methods—including those based on Differential Evolution (briefly recalled in Appendix \ref{app:DE}) and on the Optimal Transport approach introduced earlier—yields remarkably consistent results. Computational times are extremely short (on the order of one second), thereby demonstrating the robustness and efficiency of the proposed methodology.

\begin{table}[h!]
\centering
\small
\renewcommand{\arraystretch}{1.2}
\setlength{\tabcolsep}{2.2pt} 

\begin{tabular}{
  >{\raggedright\arraybackslash}p{17mm}  
  >{\centering\arraybackslash}p{31mm}
  >{\centering\arraybackslash}p{31mm}
  >{\centering\arraybackslash}p{31mm}
  >{\centering\arraybackslash}p{31mm}
  >{\centering\arraybackslash}p{31mm}
}
\toprule
\shortstack{\textbf{Parameter} \\\textbf{Metric} }&
\shortstack{\textbf{Without}\\\textbf{Constraint}\\Fixed Point Eq.\\Semi-Analytical} &
\shortstack{\textbf{Without} \\\textbf{Constraint}\\Quantization\\Diff. Evol.} &
\shortstack{\textbf{VaR 99\%}\\\textbf{Constraint}\\Quantization\\Diff. Evol.} &
\shortstack{\textbf{Without}\\\textbf{Constraint}\\Opt. Transport\\ Sinkhorn-Knopp} &
\shortstack{\textbf{VaR 99\%}\\\textbf{Constraint}\\Opt. Transport\\ Sinkhorn-Knopp} \\
\midrule
$m_1$      & 8,206,458  & 8,210,913  & 8,878,795  & 8,465,072  & 8,834,289 \\
$m_2$      & 21,452,567 & 21,453,539 & 24,397,039 & 21,971,367 & 23,925,852 \\

$p_0$      & 67.96\% & 67.99\% & 69.34\% & 68.51\% & 69.25\% \\
$p_1$      & 26.77\% & 26.74\% & 27.26\% & 26.94\% & 27.26\% \\
$p_2$      & 5.27\%  & 5.27\%  & 3.40\%  & 4.54\%  & 3.48\% \\
\midrule
VaR 99\%   & \multicolumn{5}{c}{22,531,887} \\
ES 97.5\%  & \multicolumn{5}{c}{23,636,174} \\
Worst-case & \multicolumn{5}{c}{34,616,367} \\
\bottomrule
\end{tabular}

\caption{Market risk — Fixed Point Eq. (Semi-analytical using \eqref{m1},\eqref{m2}), Quantization with Differential Evolution and Optimal Transport (Sinkhorn-Knopp) methods. Comparison with and without the VaR 99\% constraint on $m_2$.}
\label{tab:var_comparison_market_risk_compact}
\end{table}

\subsubsection{Case 2: Credit  Risk in the Banking Sector }

\setlength{\dashlinedash}{0.5pt} 
\setlength{\dashlinegap}{1.2pt}  
\setlength{\arrayrulewidth}{0.2pt} 

\begin{table}[h!]
\centering
\small
\renewcommand{\arraystretch}{1.2}
\setlength{\tabcolsep}{2.2pt}

\begin{tabular}{
  >{\raggedright\arraybackslash}p{17mm}
  >{\centering\arraybackslash}p{31mm}
  >{\centering\arraybackslash}p{31mm}
  >{\centering\arraybackslash}p{31mm}
  >{\centering\arraybackslash}p{31mm}
  >{\centering\arraybackslash}p{31mm}
}
\toprule
\shortstack{\textbf{Parameter}\\\textbf{Metric}} &
\shortstack{\textbf{Without}\\\textbf{Constraint}\\Fixed Point Eq. \\ Semi-Analytical} &
\shortstack{\textbf{Without}\\\textbf{Constraint}\\Quantization \\Diff.\ Evol.} &
\shortstack{\textbf{VaR 99.9\%}\\\textbf{Constraint}\\Quantization\\Diff.\ Evol.} &
\shortstack{\textbf{Without}\\\textbf{Constraint}\\Opt.\ Transport\\Sinkhorn-Knopp} &
\shortstack{\textbf{VaR 99.9\%}\\\textbf{Constraint}\\Opt.\ Transport\\Sinkhorn-Knopp} \\
\midrule
$m_1$ & 66,944,517 & 66,944,702 & 66,944,602 & 66,944,517 & 66,944,517 \\
$m_2$ & 455,901,571 & 455,900,234 & 455,901,519 & 455,901,571 & 455,901,571 \\
$p_0$ & 98.84\% & 98.84\% & 98.84\% & 98.84\% & 98.84\% \\
$p_1$ & 1.13\% & 1.13\% & 1.13\% & 1.13\% & 1.13\% \\
$p_2$ & 0.03\% & 0.03\% & 0.03\% & 0.03\% & 0.03\% \\

\midrule
VaR & \multicolumn{5}{c}{125,882,878} \\
ES  & \multicolumn{5}{c}{256,621,107} \\
Worst-case & \multicolumn{5}{c}{980,312,060} \\
\bottomrule
\end{tabular}

\caption{Credit risk comparison — Fixed Point Eq. (Semi-analytical using \eqref{m1},\eqref{m2}), Quantization with Differential Evolution and Optimal Transport (Sinkhorn-Knopp) methods.  Comparison with and without the VaR 99.9\% constraint on $m_2$. }
\label{tab:credit_risk_comparison_merged}
\end{table}

Within the Credit Risk framework—modeled through a Merton-type setting with a multivariate Gaussian copula—the VaR constraint on $m_2$ plays a less active role, as we can immediately see from Table \ref{tab:credit_risk_comparison_merged}. Due to portfolio concentration effects, the presence of dominant exposures shifts the loss distribution to the right, leading to a naturally high value of 
$m_2$, which already captures the fat-tail behavior of the underlying risk factors.
Moreover, due to typical combinatorial effects associated with default events, significantly different scenarios may lead to very similar loss values, thereby making risk measures structurally unstable and leading to even more fragile risk decomposition techniques, as illustrated in the Appendix \ref{AppendixCredit}.
Consequently, imposing a constraint on the value of $m_2$
 (for instance, requiring $m_2$
 to exceed the VaR) may be of limited relevance, since the tail of the distribution is already heavily skewed toward extreme values, forcing $m_2$
 into a region  that the VaR is not able to reach, even with 
the typical credit risk regulatory requirement of a 99.9\% confidence level. In other words, the extreme values of the distribution are so pronounced that the VaR no longer provides an adequate representation of tail risk, whereas $m_2$
 effectively captures the true risk embedded in the distribution—yielding substantially higher values and thereby making any additional VaR-based constraint redundant.

\begin{figure}[h!]
\centering
\includegraphics[width=1.1\textwidth]{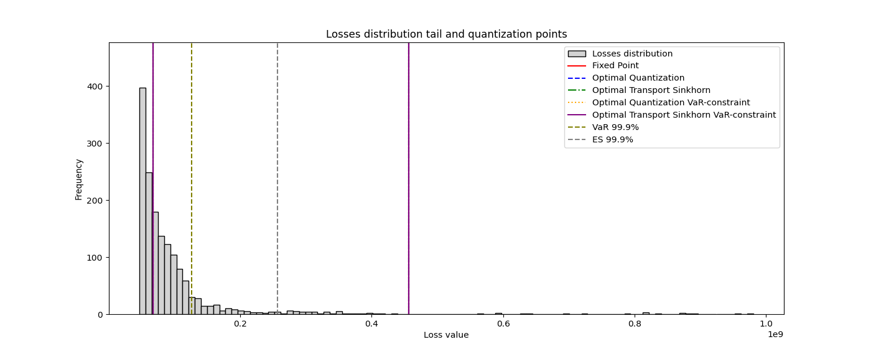} 
\caption{Credit Risk in the Banking Sector (VaR 99.9\%, ES 99.9 \%): tail of the Loss distribution.}
\label{fig:credit risk}
\end{figure}

A comparison between $m_2$
 and the worst-case scenario, together with the graphical inspection of the losses histogram (see Figure \ref{fig:credit risk}), clearly points out how 
$m_2$, acting as a barycenter of the extreme-loss scenarios, is positioned well above the VaR level, but remains consistently below the worst-case value. 
In fact, from Figure \ref{fig:credit risk} we realize a sparse distribution on the right tail, leading to extreme losses with relatively non negligible probabilities. This is well captured by a relatively high value for $m_2$ with respect to the VaR and ES that underestimate the risk.

Due to the very high magnitude and low frequency of extreme losses, the associated parameter $p_2$ has a very small value. It is worth recalling, from an intuitive standpoint and in analogy with the theory of optimal transport, that the parameters $(m,p)$ jointly minimize the “effort” required to transfer probability masses and values from a continuous distribution toward its discrete representation.
This observation underscores the inherent tendency of VaR and ES to underestimate the probability of extreme losses, as is often the case in credit risk modeling.

In conclusion, the proposed methodology once again provides a more accurate and comprehensive representation of both moderate and extreme risks than traditional measures based on VaR and ES, particularly within the complex framework of credit risk.


\subsubsection{Case 3: Market Risk in the Insurance Sector }

In the Insurance Market Risk case, the one-year horizon and 99.5\% confidence level adopted under Solvency II standards confirm the adaptability of the proposed approach to different regulatory settings, see Table \ref{tab:insurance_market_risk_comparison_merged}.

\begin{table}[h!]
\centering
\small
\renewcommand{\arraystretch}{1.2}
\setlength{\tabcolsep}{2.2pt}

\begin{tabular}{
  >{\raggedright\arraybackslash}p{17mm}
  >{\centering\arraybackslash}p{31mm}
  >{\centering\arraybackslash}p{31mm}
  >{\centering\arraybackslash}p{31mm}
  >{\centering\arraybackslash}p{31mm}
  >{\centering\arraybackslash}p{31mm}
}
\toprule
\shortstack{\textbf{Parameter}\\\textbf{Metric}} &
\shortstack{\textbf{Without}\\\textbf{Constraint}\\Fixed Point Eq. \\ Semi-Analytical} &
\shortstack{\textbf{Without}\\\textbf{Constraint}\\Quantization\\Diff.\ Evol.} &
\shortstack{\textbf{VaR 99.5\%}\\\textbf{Constraint}\\Quantization\\Diff.\ Evol.} &
\shortstack{\textbf{Without}\\\textbf{Constraint}\\Opt.\ Transport\\Sinkhorn-Knopp} &
\shortstack{\textbf{VaR 99.5\%}\\\textbf{Constraint}\\Opt.\ Transport\\Sinkhorn-Knopp} \\
\midrule
$m_1$ & 292,726,533 & 292,251,249 & 292,250,135 & 292,726,533 & 292,726,533 \\
$m_2$ & 1,326,307,230 & 1,326,305,007 & 1,326,304,534 & 1,326,307,230 & 1,326,307,230 \\
$p_0$ & 90.03\% & 90.00\% & 90.00\% & 90.03\% & 90.03\% \\
$p_1$ & 9.21\% & 9.24\% & 9.24\% & 9.21\% & 9.21\% \\
$p_2$ & 0.76\% & 0.76\% & 0.76\% & 0.76\% & 0.76\% \\
\midrule
VaR        & \multicolumn{5}{c}{1,103,006,334} \\
ES         & \multicolumn{5}{c}{1,524,216,012} \\
Worst-case & \multicolumn{5}{c}{2,791,623,043} \\
\bottomrule
\end{tabular}

\caption{Insurance market risk — Fixed Point Eq. (Semi-analytical using \eqref{m1},\eqref{m2}), Quantization with Differential Evolution and Optimal Transport (Sinkhorn-Knopp) methods.  Comparison with and without the VaR 99.5\% constraint on $m_2$.}
\label{tab:insurance_market_risk_comparison_merged}
\end{table}

\begin{figure}[h!]
\centering
\includegraphics[width=1.1\textwidth]{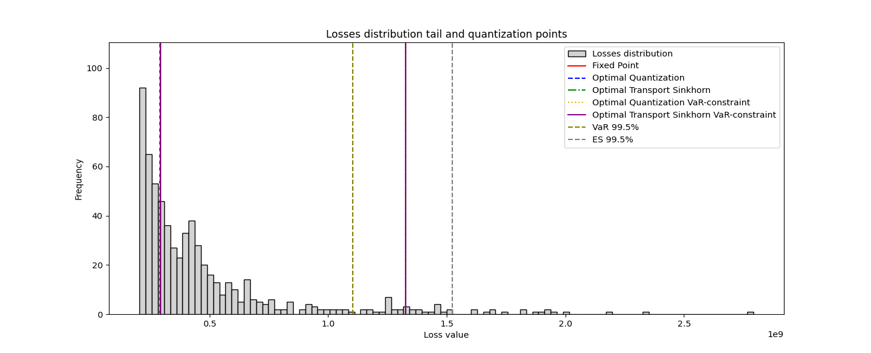} 
\caption{Insurance market risk (VaR 99.5\%, ES 99.5\%): tail of the Loss distribution.}
\label{fig:insurance market risk}
\end{figure}

In Figure \ref{fig:insurance market risk} we show the tail of the loss distribution, from which we deduce a considerable amount of losses beyond the VaR, although the difference between the parameter $m_2$
and  the VaR level is much less dramatic than in the previous case of Credit Risk. This behavior is consistent with the nature of the portfolio and the types of risks considered. Indeed, market risk within the insurance context also encompasses spread risk and migration risk (see Section \ref{SectionInsurance}). As a result, the PnL distribution may include potential jump events, yet with tail effects that are less pronounced than those typically observed in pure credit risk settings. Consequently, 
$m_2$
 assumes a meaningful position from an informative standpoint, identifying potentially significant losses, while $p_2$
—associated with extreme risk events—reflects a degree of conservatism (or prudence) broadly in line with the regulatory confidence level.

\section{Conclusion}

One of the long-standing challenges in risk management is the search for a risk measure that simultaneously satisfies the goals of objectivity, intuitiveness, and theoretical soundness. In this respect, the long and ongoing regulatory transition from Value-at-Risk (VaR) to Expected Shortfall (ES) clearly illustrates the difficulty of introducing new measures that can meet the expectations and requirements of all potential users.

In our work, we build on a frequency–severity framework, a well-established approach in the insurance domain that has recently received a rigorous theoretical foundation from  \cite{Faugeras21}, who formalized the search for optimality within this setting. Our main contribution consists in extending this framework to the three-point case, which is particularly relevant from a practical perspective, as it captures the essential structure of risk profiles through three representative outcomes: “no loss,” “small loss” (with medium-to-high probability), and “extreme loss” (with low probability). This setting is especially meaningful in the insurance field, where such discrete representations of risk are commonly employed.

The magnitude–propensity framework was tested across three distinct and relevant use cases, with the aim of assessing both the behavior of the numerical procedures and the practical implementation of the results. In the unconstrained case, the optimization is free to explore the parameter space for the optimal configuration 
$(m_1,m_2,p_1,p_2)$. We then introduced a constraint on 
$m_2$
 to align with regulatory requirements—specifically, the prescribed confidence level and time horizon associated with the Value-at-Risk (VaR) measure—thus enforcing 
$m_2\geq VaR$.

In the unconstrained setting, the numerical procedures exhibit mutual convergence and match the analytical fixed-point solution derived from the theoretical framework. This confirms that (i) the numerical algorithms are robust and reliable, and (ii) the theoretical formulation—yielding substantial computational savings—is correct. In the constrained case, where a theoretical benchmark is not yet available, the numerical methods still converge to consistent results. From a practical perspective, running multiple procedures in parallel may serve as a valuable cross-validation strategy to ensure stability and reliability of the optimal solution.

Finally, we note that the computational performance of the fixed-point approach is excellent, with execution times close to zero, confirming its efficiency and suitability for large-scale or real-time applications.

In the risk management process, an essential step is capital allocation, that is, the ex-ante assignment of the risk budget to individual business units and the ongoing monitoring to verify the risk reward performance of each business owner. This task requires rigorous methods of risk decomposition, particularly in the case of financial conglomerates or banking groups. A seminal reference in this area includes \cite{Garman97},  see also \cite{Tasche2002} and \cite{Tasche2004} for a formal rigorous framework.

As an initial approach, we consider it useful to extend the existing literature by treating the discrete variables $m_1$ and $m_2$
 as if they represented the true loss distribution, allowing the application of the above mentioned techniques. Roughly speaking, one has to calculate the expected loss of each business unit conditioned on the total loss of the parent portfolio, i.e., VaR in the classical approach, while $m_2$ in our setting. Under this perspective, risk decomposition reduces to the computation of a conditional expectation with respect to the discrete values assumed by the distribution.

We believe that our approach could be evaluated from a managerial perspective, not only as a tool for risk measurement and reporting, but also as a useful instrument to support the quantification of strategic parameters for financial institutions, such as risk appetite and risk tolerance.

\appendix

\section{Proofs}\label{appendixProofs}
\subsection{Proof of Lemma \ref{lemmaH}}
\begin{enumerate}
\item Fix $x$ such that, w.l.o.g., $h_{0}(x)<\min(h_{1}(x);h_{2}(x))$ and define
$d:=\min(h_{1}(x);h_{2}(x))-h_{0}(x)>0$. Take $0<\epsilon_{1}<d,0<\epsilon_{2}<d-\epsilon_{1}$
(that is, $\epsilon_{1}+\epsilon_{2}<d$). Since $h_{1},h_{2}$ are
continuous, also $h_{1}\wedge h_{2}$ is continuous, then there exists
$\delta>0$ such that $\forall v\in\mathbb{R}^{2}$ such that $\vert\vert v\vert\vert<\delta$
we have 
\begin{align*}
h_{0}(x+v)\leq & h_{0}(x)+\epsilon_{1},\\
h_{1}(x+v)\wedge h_{2}(x+v)\geq & h_{1}(x)\wedge h_{2}(x)-\epsilon_{2}.
\end{align*}
Now, from $h_{0}(x)=h_{1}(x)\wedge h_{2}(x)-d$ we get $h_{0}(x)+\epsilon_{1}=h_{1}(x)\wedge h_{2}(x)+\epsilon_{1}-d$,
so that 
\begin{align*}
h_{0}(x+v)\leq h_{1}(x)\wedge h_{2}(x)+\epsilon_{1}-d<h_{1}(x)\wedge h_{2}(x)-\epsilon_{2}\leq h_{1}(x+v)\wedge h_{2}(x+v).
\end{align*}
Therefore $H(x+v)=h_{0}(x+v)\quad \forall\vert\vert v\vert\vert<\delta$,
so that $H$ is differentiable, with derivative given by $\nabla H^{B}(x)(v)=\nabla H(x)v=\nabla h_{0}(x)v$. 
\item Take w.l.o.g. $h_{0}(x)=h_{1}(x)<h_{2}(x)$, so that $H(x)=h_{0}(x)\wedge h_{1}(x)$.
By repeating the same reasoning of the previous point, there exists
$\delta>0$ such that for all $\vert\vert v\vert\vert<\delta$ 
\begin{align*}
H(x+v)=h_{0}(x+v)\wedge h_{1}(x+v).
\end{align*}
Now, from Lemma 3.3 in \cite{Faugeras21} we have that $h_{0}\wedge h_{1}$
is B-differentiable, so that $H$ is also B-differentiable and $\nabla^{B}H(x)(v)=\nabla h_{1}(x)v$. 
\item We have $H(x)=h_{0}(x)=h_{1}(x)=h_{2}(x)$, so that 
\begin{align*}
H(x+v)-H(x) & =\min(h_{0}(x+v)-h_{0}(x);h_{1}(x+v)-h_{1}(x);h_{2}(x+v)-h_{2}(x))\\
 & =\min(h_{0}(x+v)-h_{0}(x);\min(h_{1}(x+v)-h_{1}(x);h_{2}(x+v)-h_{2}(x))).
\end{align*}
Now, using the inequality $\vert\min(a;b)-\min(c;d)\vert\leq\max(\vert a-c\vert;\vert b-d\vert)$
with 
\begin{align*}
a & =h_{0}(x+v)-h_{0}(x),\\
b & =\min(h_{1}(x+v)-h_{1}(x);h_{2}(x+v)-h_{2}(x)),\\
c & =\nabla h_{0}(x)v,\\
d & =\nabla^{B}(h_{1}\wedge h_{2})(x)(v),
\end{align*}
we get {\small{}
\begin{align*}
 & \vert H(x+v)-H(x)-\min(\nabla h_{0}(x)v;\nabla^{B}h_{1}\wedge h_{2})(x)(v))\vert\\
 & \leq\max(\vert h_{0}(x+v)-h_{0}(x)-\nabla h_{0}(x)v\vert;\vert\min(h_{1}(x+v)-h_{1}(x);h_{2}(x+v)-h_{2}(x))-\nabla^{B}(h_{1}\wedge h_{2})(x)v\vert)\\
 & =\max(\vert h_{0}(x+v)-h_{0}(x)-\nabla h_{0}(x)v\vert;\vert h_{1}\wedge h_{2}(x+v)-h_{1}\wedge h_{2}(x)-\nabla^{B}(h_{1}\wedge h_{2})(x)(v)\vert)\\
 & =\max(\vert\vert v\vert\vert\vert o(1)\vert;\vert h_{1}\wedge h_{2}(x+v)-h_{1}\wedge h_{2}(x)-\min(\nabla h_{1}(x)v;\nabla h_{2}(x)v)\vert)\\
 & =\vert\vert v\vert\vert\max(\vert o(1)\vert;\vert o(1)\vert),
\end{align*}
} therefore 
\begin{align*}
\nabla^{B}H(x)(v) & =\min(\nabla h_{0}(x)v;\min(\nabla h_{1}(x)v;\nabla h_{2}(x)v))\\
 & =\min(\nabla h_{0}(x)v;\nabla h_{1}(x)v;\nabla h_{2}(x)v),
\end{align*}
and the proof is complete. 
\end{enumerate}

\subsection{Proof of Theorem \ref{teoremaD}}

The proof is organized in two steps. 

\subsubsection{Step 1: B-derivatives of the function $H$}

Let $H(x_{1},x_{2})=X^{2}\wedge(X-x_{1})^{2}\wedge(X-x_{2})^{2}$
with $0<x_{1}<x_{2}$. Fix $X$ and set $h_{0}(x_{1},x_{2})=X^{2};h_{1}(x_{1},x_{2})=(X-x_{1})^{2};h_{2}(x_{1},x_{2})=(X-x_{2})^{2}$.
We have then the following seven possibilities. 

\begin{enumerate}
\item $h_{0}(x)=h_{1}(x)=h_{2}(x)$ for $x=(x_{1},x_{2})\in\{(0,0),(0,2X),(2X,0),(2X,2X)\}$.
In this case we have $H(x_{1},x_{2})=X^{2}$ and the B-derivatives
are given respectively by: 
\begin{align*}
\nabla^{B}H(0,0)(v) & =\min(0;-2Xv_{1};-2Xv_{2}),\\
\nabla^{B}H(0,2X)(v) & =\min(0;-2Xv_{1};2Xv_{2}),\\
\nabla^{B}H(2X,0)(v) & =\min(0;2Xv_{1};-2Xv_{2}),\\
\nabla^{B}H(2X,2X)(v) & =\min(0;2Xv_{1};2Xv_{2}).
\end{align*}
\item $h_{0}(x)=h_{1}(x)<h_{2}(x)$ for $x=(x_{1},x_{2})$ such that $(x_{1}=0\quad and\quad x_{2}>2X)$
or $(x_{1}=2X\quad and\quad x_{2}>2X)$. In this case we still have $H(x_{1},x_{2})=X^{2}$
and the B-derivative is given by: 
\begin{align*}
\nabla^{B}H(0,x_{2})(v) & =\min(0;-2Xv_{1}),\\
\nabla^{B}H(2X,x_{2})(v) & =\min(0;2Xv_{1}),
\end{align*}
provided that $x_{2}>2X$.
\item $h_{0}(x)=h_{2}(x)<h_{1}(x)$ for $(x_{1}>2X\quad and\quad x_{2}=0)$ or $(x_{1}>2X\quad and\quad x_{2}=2X)$.
In this case we still have $H(x_{1},x_{2})=X^{2}$ and the B-derivative
is given by: 
\begin{align*}
\nabla^{B}H(x_{1},0)(v) & =\min(0;-2Xv_{2}),\\
\nabla^{B}H(x_{1},2X)(v) & =\min(0;2Xv_{2}),
\end{align*}
provided that $x_{1}>2X$. 
\item $h_{1}(x)=h_{2}(x)<h_{0}(x)$ for $(x_{1}+x_{2}=2X\quad and\quad 0<x_{1}<x_{2}<2X)$.
This time we have $H(x_{1},x_{2})=(X-x_{1})^{2}=(X-x_{2})^{2}$ and
the B-derivative is given by: 
\begin{align*}
\nabla^{B}H(x_{1},2X-x_{1})(v) & =\min(-2(X-x_{1})v_{1};-2(X-x_{2})v_{2}),
\end{align*}
provided that $x_{1}+x_{2}=2X$ and $0<x_{1}<x_{2}<2X$. 
\item $h_{0}(x)<h_{1}(x)\wedge h_{2}(x)$ for $x_{1}>2X$. We have $H(x_{1},x_{2})=X^{2}$
and the B-derivative is zero. 
\item $h_{1}(x)<h_{0}(x)<h_{2}(x)$ for $(x_{1}<2X\quad and\quad x_{1}+x_{2}>2X)$.
This time we have $H(x_{1},x_{2})=(X-x_{1})^{2}$ and the B-derivative
is given by: 
\begin{align*}
\nabla^{B}H(x_{1},x_{2})(v) & =-2(X-x_{1})v_{1},
\end{align*}
provided that $x_{1}<2X$ and $x_{1}+x_{2}>2X$. 
\item $h_{2}(x)<h_{0}(x)<h_{1}(x)$ for $(x_{1}+x_{2}<2X)$. This time we
have $H(x_{1},x_{2})=(X-x_{2})^{2}$ and the B-derivative is given
by: 
\begin{align*}
\nabla^{B}H(x_{1},x_{2})(v) & =-2(X-x_{2})v_{2},
\end{align*}
provided that $x_{1}+x_{2}<2X$. 
\end{enumerate}

\subsubsection{Step 2: B-differentiability of the objective function $D$ and critical points for the distortion}

Set 
\begin{align}
\nabla^{B}D(x)(v):=\mathbb{E}[\nabla^{B}H(x)(v)],
\end{align}
and consider now the different possibilities listed in the previous
subsection.

\begin{enumerate}
\item The point $x=(0,0)$ is critical for the distortion function if $\nabla^{B}D(0,0)(v)\geq0\forall v\in\mathbb{R}^{2}$.
Now, 
\begin{align*}
\nabla^{B}D(0,0)(v) & =\mathbb{E}[\min(0;-2Xv_{1};-2Xv_{2})]\\
 & =\mathbb{E}[\min(-2Xv_{1};-2Xv_{2})]<0
\end{align*}
for $v_{1}>0,v_{2}>0$, therefore the point $x=(0,0)$ is not critical
for the distortion. The same holds true for $(0,2X)$, since $\nabla^{B}D(0,2X)(v)=\mathbb{E}[\min(0;-2Xv_{1};-2Xv_{2})]<0$
for $v_{1}>0,v_{2}<0$. Using the same argument one can exclude also
$(2X,0),(2X,2X)$. 
\item Consider now $x_{2}>2X$, we get 
\begin{align*}
\nabla^{B}D(0,x_{2})(v) & =\mathbb{E}[\min(0;-2Xv_{1})]<0\quad \quad for\quad v_{1}>0,\\
\nabla^{B}D(2X,x_{2})(v) & =\mathbb{E}[\min(0;2Xv_{1})]<0\quad\quad \quad for\quad v_{1}<0,
\end{align*}
so that the $(0,x_{2}),(2X,x_{2})$ are not critical for $x_{2}>2X$.
\item Let $x_{1}>2X$, 
\begin{align*}
\nabla^{B}D(x_{1},0)(v) & =\mathbb{E}[\min(0;-2Xv_{2})]<0\quad \quad for\quad v_{2}>0,\\
\nabla^{B}D(x_{1},2X)(v) & =\mathbb{E}[\min(0;2Xv_{2})]<0\quad \quad for\quad v_{2}<0,
\end{align*}
therefore $(x_{1},0),(x_{1},2X)$ are not critical for $x_{1}>2X$.
\item Consider $x_{1}+x_{2}=2X$ and $0<x_{1}<x_{2}<2X$: 
\begin{align*}
\nabla^{B}D(x_{1},2X-x_{1})(v) & =\mathbb{E}[\min(-2(X-x_{1})v_{1};-2(X-x_{2})v_{2})]<0\quad \quad for\quad v_{1}>0,v_{2}<0,
\end{align*}
therefore $(x_{1},2X-x_{1})$ is not critical for $x_{1}+x_{2}=2X$
and $0<x_{1}<x_{2}<2X$. 
\item We have $\nabla^{B}D(x_{1},x_{2})(v)=0$ for $x_{1}>2X$. 
\item Take $x_{1}<2X$ and $x_{1}+x_{2}>2X$: 
\begin{align*}
\nabla^{B}D(x_{1},x_{2})(v)= & \mathbb{E}[-2(X-x_{1})v_{1}{\bf 1}_{x_{1}<2X}{\bf 1}_{x_{1}+x_{2}>2X}]\\
 & +\mathbb{E}[\min(0;2Xv_{1}){\bf 1}_{x_{1}=2X}{\bf 1}_{x_{2}>2X}]\\
 & +\mathbb{E}[\min(-2(X-x_{1})v_{1};-2(X-x_{1})v_{2}){\bf 1}_{x_{1}<2X}{\bf 1}_{x_{1}+x_{2}=2X}],
\end{align*}
so that $(x_{1},x_{2})$ may be a critical point. 
\item Finally, let consider $x_{1}+x_{2}<2X$: 
\begin{align*}
\nabla^{B}D(x_{1},x_{2})(v) & =\mathbb{E}[-2(X-x_{2})v_{2}{\bf 1}_{x_{1}+x_{2}<2X}],
\end{align*}
then $(x_{1},x_{2})$ may be a critical point. 
\end{enumerate}
In conclusion, it turns out that the function $v\rightarrow\nabla^{B}D(x)(v)$
is positively homogeneous for $x\in\mathbb{R}_{+}^{2}$, hence $D$
is B-differentiable. Moreover, collecting the different cases one gets
\eqref{BD}, and the proof of the theorem is complete.


\section{Intrinsic Instability of the Quantile in Credit Risk Modeling}\label{AppendixCredit}

Extreme quantiles such as the 99.9\% VaR play a central role in the computation of the 
\textit{Default Risk Charge} (DRC).  
However, when applied to concentrated credit portfolios, these percentiles may exhibit a 
structural instability that does not originate from Monte Carlo error, 
but from the intrinsic combinatorial geometry of the loss distribution.  
Before analysing this phenomenon, we introduce the simplified analytical setup used to isolate and quantify such behavior.

\medskip
\noindent\textit{Analytical setup.}
Consider a reduced set of the $N$ largest obligors (covering the majority of portfolio EAD) 
and model defaults via a one–factor Gaussian copula.  
For obligor $n=1,\ldots,N$, we define
\[
Z_n=\sqrt{\rho_n}\,U + \sqrt{1-\rho_n}\,\varepsilon_n,
\]
where $U,\varepsilon_n$ are independent standard Gaussian random variables, and a default occurs whenever $Z_n$ falls below the threshold given by the corresponding default probability, i.e. 
$b_n:=\Phi^{-1}(\mathrm{PD}_n)$.  
The portfolio loss in 
\ref{DRC}
is therefore
\begin{equation*}
Loss_{DRC}
= \sum_{n=1}^{N}
EAD_n\cdot \mathbf{1}_{\{Z_n \le b_n\}}\cdot LGD_n,
\label{eq:LossPTF}
\end{equation*}
with deterministic LGD in this experiment.  

Since each obligor may be in default or not, the loss distribution is supported on the 
$2^{N}$ possible default/non–default combinations.  
For each configuration $r\subset\{1,\ldots,N\}$, the corresponding loss is
\begin{equation*}
\mathrm{Loss}_r = \sum_{n\in r}EAD_n\cdot
\mathbf{1}_{\{Z_n \le b_n\}}\cdot LGD_n,
\end{equation*}
while its probability under the copula reduces to a one–dimensional integral:
\begin{equation}
P_r = 
\int_{-\infty}^{+\infty}
\left[
    \prod_{n\in r}
    \Phi\!\Big(\tfrac{b_n-\sqrt{\rho_n}u}{\sqrt{1-\rho_n}}\Big)
\right]
\left[
    \prod_{n\notin r}
    \Big(1-\Phi\!\Big(\tfrac{b_n-\sqrt{\rho_n}u}{\sqrt{1-\rho_n}}\Big)\Big)
\right]
\phi(u)\,du ,
\label{eq:Pr}
\end{equation}
evaluated via Gaussian quadrature.  
Sorting all $(\mathrm{Loss}_r, P_r)$ pairs yields the exact loss distribution and, in particular,
the exact 99.9\% Default Risk Charge DRC is defined as
\begin{equation*}
\mathrm{DRC}
= \min \big\{\mathrm{Loss}_r : P_{\mathrm{cum}}(r)\ge 0.999 \big\},
\end{equation*}
where $P_{\mathrm{cum}}(r)=\sum_{\vert m\vert \le \vert r\vert }P_m$, where the sum is extended to uncompatible events.

\medskip
\noindent\textit{Focusing on the percentile instability.}
To analyse how the percentile behaves in the tail,  
we examine the symmetric window 
\[
\{k-500,\ldots,k+500\},
\]
where $k$ is the index of the order statistic corresponding to the 99.9\% quantile.  
This window contains the ranked scenarios whose cumulative probabilities lie immediately below and above 0.999—precisely where small perturbations may change the identity of the selected quantile.

\begin{figure}[h!]
\centering
\makebox[\textwidth][c]{\includegraphics[width=1.20\textwidth]{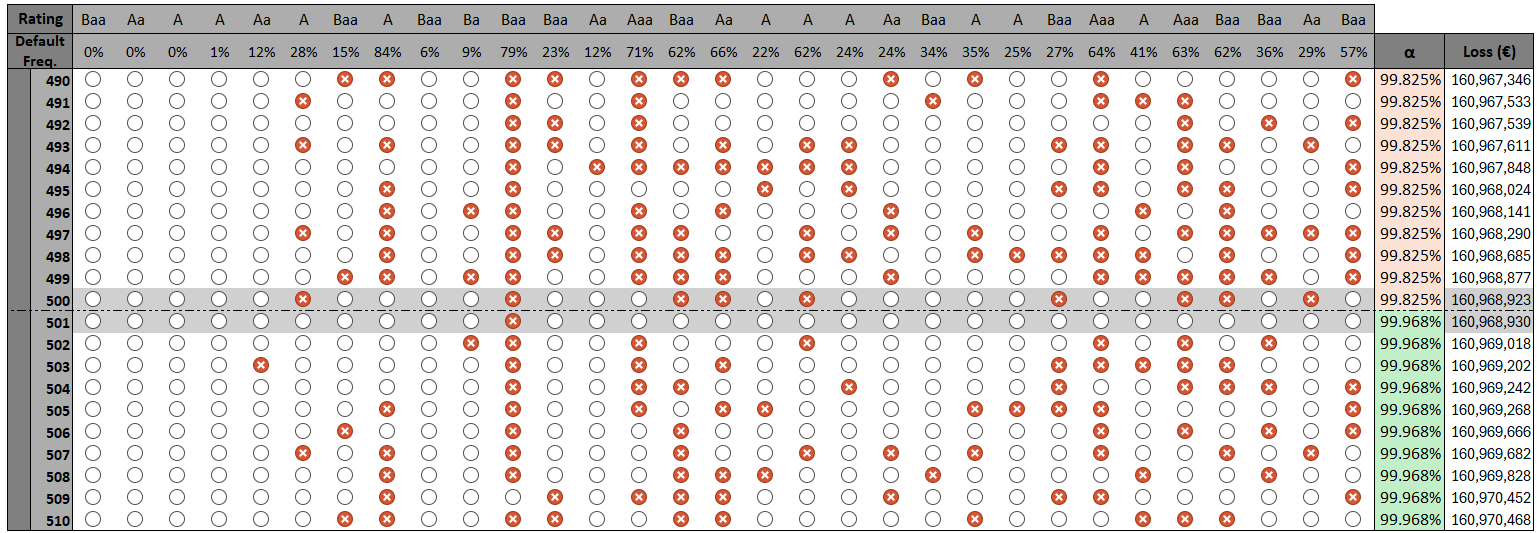}}
\caption{Tail–stability analysis around the 99.9\% VaR (DRC). The figure shows the ordered loss contributions in the neighbourhood of the 99.9\% percentile, highlighting how consecutive scenarios exhibit nearly identical loss levels. 
Because cumulative probabilities are extremely small in the tail, even minimal perturbations in PDs, correlations, exposures or in the ordering of almost equal losses may shift the DRC from $\mathrm{Loss}^{(k)}$ to $\mathrm{Loss}^{(k+1)}$ without any material change in magnitude. 
The visual alignment of points illustrates this intrinsic sensitivity of the DRC to local fluctuations in the extreme tail.
}
\label{fig:Default_scenarios_n31 v7}
\end{figure}

Figure~\ref{fig:Default_scenarios_n31 v7} displays, for each scenario in the quantile window,  
(i) the default pattern across the $N=31$ obligors (orange dots),  
(ii) the cumulative probability $\alpha_r$,  
and (iii) the corresponding loss $\mathrm{Loss}_r$.  
The figure highlights a crucial structural phenomenon:  
\textit{despite the fact that default patterns differ significantly across scenarios, the resulting losses in this tail region are numerically almost identical}. 
In other words, the instability is note related to the  magnitude of the loss, but rather to the specific configuration of the defaulting obligors.  
This arises because many combinations of $\mathrm{EAD}_n\mathrm{LGD}_n$ sum to nearly the same value, an effect amplified by the combinatorial explosion of the support.

Consequently, since adjacent cumulative levels satisfy
\[
\alpha_{r+1}-\alpha_r = P_{r+1},
\]
and the probabilities $P_r$ are extremely small near the tail, a minimal perturbation 
(e.g.\ slight changes in PDs, correlations, exposures, or simply a reordering of nearly equal losses)
may shift the DRC from $\mathrm{Loss}^{(k)}$ to $\mathrm{Loss}^{(k\pm 1)}$ without any meaningful variation in loss magnitude.  
The percentile therefore experiences a discontinuous jump driven purely by ranking, not by economic information.

\medskip
\noindent\textit{Implication for VaR-based DRC.}
The analysis makes clear that, for concentrated credit portfolios and extreme confidence levels,
the usability of VaR as a tail-risk measure is inherently limited: the 99.9\% percentile becomes 
highly sensitive to microscopic changes in the loss distribution and does not reliably reflect  
the underlying risk structure.  
This motivates the use of tail-aggregation methods—such as the magnitude–propensity framework introduced in our paper—which summarize information across multiple extreme-loss configurations rather than relying on a single order statistic.

\section{The Sinkhorn--Knopp (SK) Algorithm}\label{appendixOT}

In this appendix,  we provide some details on the numerical procedure adopted for the solution of the entropically regularized optimal transport problem, following the classical matrix-scaling approach originally introduced in \cite{Sinkhorn1964,SinkhornKnopp1967} and later applied to optimal transport in \cite{Cuturi2013}.

Consider two discrete probability measures 
$\mu = \sum_{i=1}^{n} p_i \, \delta_{x_i}$ and 
$\nu = \sum_{j=1}^{m} q_j \, \delta_{y_j}$, where
$\bm p = (p_1,\dots,p_n)^\top$ and $\bm q = (q_1,\dots,q_m)^\top$
are probability vectors, i.e.\ $p_i \ge 0$, $q_j \ge 0$ and
$\sum_i p_i = \sum_j q_j = 1$.
Let $C \in \mathbb{R}^{n \times m}$ be the cost matrix with entries
$c_{ij} = c(x_i,y_j)$ for a given cost function $c(\cdot,\cdot)$.
The entropically regularized optimal transport problem reads
\begin{equation}
  \min_{\Pi \in \mathbb{R}_+^{n \times m}}
  \left\{
    \langle C,\Pi\rangle 
    + \varepsilon \sum_{i=1}^{n} \sum_{j=1}^{m}
       \Pi_{ij}\bigl(\log \Pi_{ij} - 1\bigr)
  \right\}
  \quad \text{s.t.} \quad
  \Pi \bm 1_m = \bm p, \;
  \Pi^\top \bm 1_n = \bm q ,
  \label{eq:OT_entropic}
\end{equation}
where $\varepsilon > 0$ is the regularization parameter,
$\langle C,\Pi \rangle = \sum_{ij} c_{ij}\Pi_{ij}$,
and $\bm 1_k$ denotes the $k$-dimensional vector of ones.

Introducing the kernel matrix
\begin{equation*}
  K_{ij} = \exp\!\left(-\frac{c_{ij}}{\varepsilon}\right), 
  \qquad i=1,\dots,n,\; j=1,\dots,m,
\end{equation*}
it is well known that any solution of \eqref{eq:OT_entropic} can be written in the
scaling form
\begin{equation}
  \Pi^\star = \operatorname{diag}(\bm u)\, K \,\operatorname{diag}(\bm v),
  \label{eq:pi_scaling}
\end{equation}
for some positive vectors 
$\bm u \in \mathbb{R}^n_{++}$ and $\bm v \in \mathbb{R}^m_{++}$.
The Sinkhorn--Knopp algorithm is an iterative matrix scaling procedure
that computes $(\bm u,\bm v)$ such that the marginal constraints
in \eqref{eq:OT_entropic} are satisfied.

Let $\bm u^{(0)} \in \mathbb{R}^n_{++}$ and $\bm v^{(0)} \in \mathbb{R}^m_{++}$
be initial scaling vectors, typically chosen as vectors of ones.
At iteration $\ell = 0,1,2,\dots$, the algorithm performs the updates
\begin{align}
  u^{(\ell+1)}_i 
  &= \frac{p_i}{\sum_{j=1}^{m} K_{ij} v^{(\ell)}_j},
     \qquad i=1,\dots,n, \label{eq:sinkhorn_u}\\
  v^{(\ell+1)}_j 
  &= \frac{q_j}{\sum_{i=1}^{n} K_{ij} u^{(\ell+1)}_i},
     \qquad j=1,\dots,m
    \label{eq:sinkhorn_v}
\end{align}

The sequence $\{(\bm u^{(\ell)},\bm v^{(\ell)})\}_{\ell \ge 0}$
generated by \eqref{eq:sinkhorn_u}--\eqref{eq:sinkhorn_v} defines a sequence of transport
plans
\begin{equation*}
  \Pi^{(\ell)} 
  = \operatorname{diag}\bigl(\bm u^{(\ell)}\bigr)\,
    K \,
    \operatorname{diag}\bigl(\bm v^{(\ell)}\bigr),
\end{equation*}
whose row and column sums converge to $\bm p$ and $\bm q$, respectively, under mild
assumptions on $K$ (e.g.\ all entries strictly positive).
In practice, the iterations are stopped when the violation of the marginals
is below a prescribed tolerance $\tau > 0$, i.e.\ when
\begin{equation}
  \left\| \Pi^{(\ell)} \bm 1_m - \bm p \right\|_1
  + \left\| (\Pi^{(\ell)})^\top \bm 1_n - \bm q \right\|_1
  \le \tau.
  \label{eq:sk_tolerance}
\end{equation}

For numerical stability, in particular when $\varepsilon$ is small
and the entries of $K$ may underflow, the updates
\eqref{eq:sinkhorn_u}--\eqref{eq:sinkhorn_v} can be implemented in the log-domain.
Defining $\bm \alpha^{(\ell)} = \log \bm u^{(\ell)}$ and
$\bm \beta^{(\ell)} = \log \bm v^{(\ell)}$, one replaces the matrix–vector
products by log-sum-exp operations, which reduces the risk of numerical
underflow/overflow while leaving the fixed point \eqref{eq:pi_scaling} unchanged.

Once convergence has been reached (according to \eqref{eq:sk_tolerance}),
the transport plan $\Pi^\star$ in \eqref{eq:pi_scaling} is used to compute
the (regularized) optimal transport cost
\begin{equation*}
  \mathcal{W}_\varepsilon(\bm p,\bm q)
  = \langle C, \Pi^\star \rangle
    + \varepsilon \sum_{i,j}
      \Pi^\star_{ij} \bigl(\log \Pi^\star_{ij} - 1\bigr),
\end{equation*}
which is then employed in the subsequent stages of the analysis.

\section{The Differential Evolution (DE) Algorithm}\label{app:DE}

In this appendix, we provide some details on the numerical procedure we adopted for the optimization problem.

The Differential Evolution (DE) algorithm, originally proposed by 
\cite{StornPrice1997}, is a stochastic and population-based global optimization technique belonging 
to the family of evolutionary algorithms. It is particularly effective for solving 
non-convex, non-differentiable, and multi-modal optimization problems. 
Unlike gradient-based methods, DE relies exclusively on function evaluations, which makes it 
robust even when the objective function $f(\mathbf{x})$ is noisy or discontinuous.

The algorithm evolves a population of candidate solutions across generations through a 
sequence of variation mechanisms — \textit{mutation}, \textit{crossover}, and \textit{selection}. 
At each iteration, these operations allow the population to progressively explore the search 
space and concentrate around high-performing regions, eventually converging to a near-optimal 
solution for the problem under consideration.
\medskip
\\
The optimization problem is defined as
\begin{equation}
    \min_{\mathbf{x} \in \Omega} f(\mathbf{x}),
    \label{eq:DE_min_problem}
\end{equation}
where $\Omega \subseteq \mathbb{R}^d$ denotes the feasible domain of the
$d$-dimensional decision vector $\mathbf{x}$.
The approach consists in evolving a set of candidate solutions over successive generations.

At generation $g$, the algorithm maintains a population of $N_P$ candidate
solutions,
\begin{equation}
    \mathbf{x}_{i,g} = \bigl(x_{i,g}^{(1)}, x_{i,g}^{(2)}, \ldots,
    x_{i,g}^{(d)}\bigr), 
    \qquad i = 1,2,\ldots,N_P,
    \label{eq:DE_population}
\end{equation}
which evolve iteratively through mutation, crossover, and selection.
\medskip
\\
\textit{Mutation.} For each target vector $\mathbf{x}_{i,g}$, a mutant vector
$\mathbf{v}_{i,g}$ is generated according to the classical
\texttt{DE/rand/1/bin} scheme,
\begin{equation}
    \mathbf{v}_{i,g} = \mathbf{x}_{r_1,g} 
    + F \bigl( \mathbf{x}_{r_2,g} - \mathbf{x}_{r_3,g} \bigr),
    \label{eq:DE_mutation}
\end{equation}
where $r_1, r_2, r_3 \in \{1,\ldots,N_P\}$ are distinct indices different from
$i$, and $F \in [0,2]$ is the mutation factor controlling the amplification of
the differential variation.
Equation \eqref{eq:DE_mutation} therefore perturbs a base vector
$\mathbf{x}_{r_1,g}$ by adding a scaled difference of two other individuals,
thereby promoting exploration of the search space.
\medskip
\\
\textit{Crossover.} To increase population diversity, a trial vector
$\mathbf{u}_{i,g}$ is formed by combining components of the mutant and target
vectors:
\begin{equation}
    u_{i,g}^{(j)} =
    \begin{cases}
        v_{i,g}^{(j)}, &
        \text{if } \mathrm{rand}_j(0,1) \le C_R
        \text{ or } j = j_{\mathrm{rand}},\\[2mm]
        x_{i,g}^{(j)}, & \text{otherwise},
    \end{cases}
    \qquad j = 1,\ldots,d,
    \label{eq:DE_crossover}
\end{equation}
where $C_R \in [0,1]$ is the crossover probability and $j_{\mathrm{rand}}$ is a
randomly chosen index ensuring that at least one component originates from the
mutant vector.
Hence, \eqref{eq:DE_crossover} mixes the information contained in
$\mathbf{x}_{i,g}$ and $\mathbf{v}_{i,g}$, with the parameter $C_R$ regulating
the expected proportion of mutated components.
\medskip 
\\
\textit{Selection.} Selection is then performed to determine whether the trial
vector replaces the target vector in the next generation:
\begin{equation}
    \mathbf{x}_{i,g+1} =
    \begin{cases}
        \mathbf{u}_{i,g}, & \text{if } f(\mathbf{u}_{i,g}) \le f(\mathbf{x}_{i,g}),\\[1mm]
        \mathbf{x}_{i,g}, & \text{otherwise.}
    \end{cases}
    \label{eq:DE_selection}
\end{equation}
According to \eqref{eq:DE_selection}, the better (or equal) solution survives,
which guarantees that the population’s overall fitness does not deteriorate
over successive generations.  
The evolutionary process continues until a stopping criterion is met, such as a
maximum number of generations $G_{\max}$, or when the relative improvement in
the best fitness value falls below a threshold $\varepsilon > 0$, i.e.,
\begin{equation}
    \frac{\bigl|f_{\mathrm{best},g} - f_{\mathrm{best},g-1}\bigr|}
         {\bigl|f_{\mathrm{best},g-1}\bigr|}
    < \varepsilon.
    \label{eq:DE_stopping}
\end{equation}
Condition \eqref{eq:DE_stopping} prevents unnecessary iterations once the
algorithm has essentially converged.

\medskip

To apply the approach to empirical data in our context, the algorithm is used
to solve the nonlinear systems defining the quantities
$(m_x,p_x)$ and $(m_{1},m_{2},p_{1},p_{2})$ for the 2-point and 3-point cases,
respectively. In a historical simulation setting, the vector of the $S$ P\&Ls is
interpreted as the empirical distribution of the underlying risk factor.
Each point in the vector is therefore assigned the probability mass
\begin{equation}
    p_s = \frac{1}{S}, \qquad s = 1,\ldots,S,
    \label{eq:empirical_mass}
\end{equation}
with $S = 250$ or $S = 500$ in most of our applications.

Once the P\&Ls have been reordered, we denote by $\mathrm{PnL}(s)$ the
empirical cumulative distribution, which can be written as
\begin{equation}
    F\bigl(\mathrm{PnL}(s)\bigr) = \frac{s}{S},
    \qquad s = 1,\ldots,S,
    \label{eq:empirical_cdf_grid}
\end{equation}
while for any generic point $z \in \mathbb{R}$ the empirical distribution
function is
\begin{equation}
    F(z) = \frac{\#\{s : \mathrm{PnL}_s \le z\}}{S}.
    \label{eq:empirical_cdf_generic}
\end{equation}
Equations \eqref{eq:empirical_mass}–\eqref{eq:empirical_cdf_generic} provide
the empirical probabilities and cumulative distribution that constitute the
input to the objective function optimized by the DE algorithm in
\eqref{eq:DE_min_problem}.

\bibliography{biblio}
\bibliographystyle{apalike}

\end{document}